\documentclass{aa}

\usepackage{epsfig,graphicx,lscape,longtable}
\begin{document}

\title{The VLT-FLAMES survey of massive stars: constraints on stellar
evolution from the chemical
compositions of rapidly rotating Galactic and Magellanic Cloud B-type 
stars\thanks{Based on observations at the European Southern
Observatory in programmes 171.0237 and 073.0234}} 

\author{I. Hunter\inst{1} \and I. Brott\inst{2} \and N. Langer\inst{2} 
\and D.J. Lennon\inst{3} \and P.L. Dufton\inst{1} \and I.D. Howarth\inst{4} 
\and R.S.I. Ryans\inst{1} \and C. Trundle\inst{1} \and 
C.J. Evans\inst{5} \and A. de Koter\inst{6,2} \and 
S.J. Smartt\inst{1}}

\institute{Astrophysics Research Centre, School of Mathematics \& Physics, The
   Queen's University of Belfast, Belfast, BT7 1NN, Northern Ireland, UK 
\and Astronomical Institute, Utrecht University, Princetonplein 5,
NL-3584CC, Utrecht, Netherlands
\and Space Telescope Science Institute, 3700 San Martin Drive, Baltimore, 
MD 21218, USA
\and Department of Physics and Astronomy, University College London, Gower 
Street, London WC1E 6BT, UK
\and UK Astronomy Technology Centre, Royal Observatory, Edinburgh, 
Blackford Hill, Edinburgh, EH9 3HJ, UK
\and Astronomical Institute Anton Pannekoek, University of Amsterdam, 
Kruislaan 403, 1098 SJ Amsterdam, Netherlands}
       
\offprints{I. Hunter,\\ \email{I.Hunter@qub.ac.uk}}

\date{Received; accepted }

\abstract{}{
We have previously analysed the spectra of 135 early B-type stars in the 
Large Magellanic Cloud (LMC) and found several groups of stars that have 
chemical compositions that conflict with the theory of rotational 
mixing. Here we extend this study to Galactic and Small Magellanic
Cloud (SMC) metallicities.}
{We provide chemical compositions for $\sim$50 Galactic and $\sim$100 SMC 
early B-type stars and compare these to the LMC results.
These samples cover a range of projected rotational velocities up to
$\sim$300\,km\,s$^{-1}$ and hence are well suited to testing rotational
mixing models. The surface nitrogen abundances are utilised as a probe of
the mixing process since nitrogen is synthesized in the core of the stars 
and mixed to the surface.}
{In the SMC, we find a population of slowly rotating nitrogen-rich stars 
amongst the early B~type core-hydrogen burning stars, which is comparable to
that found previously in the LMC. The identification of non-enriched rapid 
rotators in the SMC is not possible due to the relatively high upper limits 
on the nitrogen abundance for the fast rotators. In the Galactic sample we 
find no significant enrichment amongst the core hydrogen-burning stars,
which appears to be in contrast with the expectation from both rotating 
single-star and close binary evolution models. 
However, only a small number of the rapidly rotating stars have evolved
enough to produce a significant nitrogen enrichment, and these may be analogous 
to the non-enriched rapid rotators previously found in the LMC sample.
Finally, in each metallicity regime, a population of highly enriched 
supergiants is observed, which cannot be the immediate descendants of 
core-hydrogen burning stars. Their abundances are, however, compatible with 
them having gone through a previous red supergiant phase. Together,
these observations paint a complex picture of the nitrogen enrichment 
in massive main sequence and supergiant stellar atmospheres, where age
and binarity cause crucial effects.
Whether rotational mixing is required to understand our results remains 
an open question at this time, but could be answered by identifying 
the true binary fraction in those groups of stars that do not agree 
with single-star evolutionary models.}{}
   
\titlerunning{Chemical compositions of B-type stars}

\maketitle

\section{Introduction} \label{s_intro}

Rotation is generally considered to be of critical importance for theoretical
models of massive star evolution and, specifically, rotationally-induced mixing,
where material is mixed from the stellar core into the photosphere (Heger \&
Langer \cite{heg00a}; Meynet \& Maeder  \cite{mey00}). For example, rotation has
been used as an explanation of blue to red supergiant ratios (Maeder \& Meynet
\cite{mae01}) and of Wolf-Rayet populations as a function of metallicity (Meynet
\& Maeder \cite{mey05}; Vink \& de Koter \cite{vin05}). Additionally, for a
single massive star to end its life with an associated long gamma-ray burst,
chemically homogeneous evolution through rapid rotation may be required (Yoon \& Langer
\cite{yoo05}; Woosley \& Heger \cite{woo06}). 

Stellar evolution models including rotation predict a surface enrichment 
of helium and nitrogen 
with an associated carbon and oxygen depletion during the main sequence evolution. 
Massive star surface 
abundance anomalies (particularly nitrogen) have long been observed, by, 
for example, Walborn (\cite{wal70}), Dufton (\cite{duf72}) and, more 
recently, Gies \& Lambert (\cite{gie92}), Kilian (\cite{kil92}), 
Bouret et al. (\cite{bou03}), Dufton et al. (\cite{duf05}),
Lennon et al. (\cite{len03}), Trundle \& Lennon (\cite{tru05}), 
Korn et al. (\cite{kor02}) and Venn (\cite{ven99}).
However, the majority of these analyses have focused on one of
two groups: narrow lined 
main-sequence stars (i.e. stars with low projected rotational velocities) 
or blue supergiants. For example, Kilian considered 21 main sequence Galactic
objects all with projected rotational velocities of less than 63 km\,s$^{-1}$.

The magnitude of the predicted mixing has generally been compared to and
calibrated against these two groups. It follows from the stellar evolution 
models that the more rapidly a star rotates, the more mixing will occur, and
hence the greater the nitrogen (and helium) surface enrichment that should 
be observed. Unfortunately spectroscopic studies of rapidly rotating stars 
are relatively rare. Lennon et al. (\cite{len91}) studied the O9.5V star 
HD\,93521 with a projected rotational velocity of approximately 400 km\,s$^{-1}$.
They deduced an enhanced helium abundance that they ascribed to rotational 
mixing. Howarth and Smith (\cite{how01}) analysed spectra of
three rapidly rotating O-type stars (including HD\,93521) using models that 
allowed for variations in temperature
and gravity across the stellar surface and again found evidence for helium
enhancements. By contrast, Villamariz and Herrero (\cite{vil05}) obtained 
a normal helium abundance together with a nitrogen enrichment for one of the 
stars ($\zeta$\, Oph) discussed by Howarth and Smith. 

Vrancken et al. (\cite{vra97}) analysed the spectra of two B-type targets 
in NGC\,2244, with projected rotational velocities between 200 and 300 
km\,s$^{-1}$. The analysis was undertaken differentially with respect to a
cluster member with a small projected rotational velocity. No significant 
abundance anomalies were identified with indeed the two fast rotators being found 
to have a small {\em underabundance} of nitrogen. Daflon et al. (\cite{daf01})
analysed twelve Galactic stars in Cygnus associations with projected rotational
velocities between 60 and 150 km\,s$^{-1}$. On average they had subsolar
abundances that agreed with those found for a sample of eight stars
with very low projected rotational velocities. However two targets (with 
projected rotational velocities of 100 and 142 km\,s$^{-1}$) showed relative
nitrogen enhancements of nearly a factor of two. Korn et al. (\cite{kor05})
analysed three LMC targets with projected rotational velocities of 
approximately 130 km\,s$^{-1}$\ but again found no evidence for either enhanced 
helium or nitrogen abundances.

The measurement of surface abundances of a large sample of core-hydrogen
burning, rapidly rotating, massive stars is clearly necessary for both testing and
calibrating the mixing theory and this was one of the primary drivers of
the VLT-FLAMES survey of massive stars. This Large Program on the VLT
with the FLAMES instrument 
(PI.: S.J. Smartt) focused on OB-type stars in the Galaxy and Magellanic Clouds
(Evans et al. \cite{eva05}; Evans et al. \cite{eva06}, hereafter Paper I and Paper II).
Over 700 O- and B-type stars were observed across these three metallicity 
regimes. Both the large number of objects and the different
metallicity regimes allow many of the theoretical predictions of evolutionary models to
be tested.

The principle outcomes of this survey related to rotation are as follows. 
Mokiem et al. (\cite{mok06}, \cite{mok07}) have analysed the O-type stars
in the sample ($\sim$50 objects) and used their mass-loss rates to derive
the wind-momentum luminosity relation. They show that at lower metallicity
stars rotate faster since they have lower mass loss rates. Dufton et al.
(\cite{duf06}, hereafter Paper III) have shown that Galactic cluster stars
(observed in the survey) rotate significantly faster than stars in the
Galactic field. This was consistent with studies  of stars in the double
cluster h and $\chi$\ Persei (Slettebak \cite{sle68};  Strom et al.
\cite{str05}) and other clusters (Wolff et al. \cite{wol07}; Huang and Gies
\cite{hua06a}). Hunter et al. (\cite{hun08a}, hereafter Paper IV) have
derived rotational velocities for the Magellanic Cloud B-type stars from
the survey and show that stars at low metallicity rotate faster than in
higher metallicity regimes. For two clusters (NGC\,330 and NGC\,2004),
these results complement the recent studies of Martayan et al. (\cite{mar06,
mar07}), who also utilised the FLAMES spectrograph albeit at  lower spectral
resolution. Additionally in Paper IV, it was suggested that the observed
population of B-type supergiants cannot be explained by  normal single star
evolution and either binarity or blue-loops needed to be invoked to model
the population.

Hunter et al. (\cite{hun07}, hereafter Paper V) have performed a detailed
chemical composition analysis of approximately 50 narrow-lined B-type stars from
the survey and utilised these objects to estimate the baseline chemical
compositions of the Magellanic Clouds, thereby  complementing previous studies
mainly of H\,{\sc II} regions (summarized, for  example, by Garnett \cite{gar99}).
Additionally as found from H\,II region analyses and other more limited stellar
samples (for example, Korn et al. \cite{kor02}), they confirmed that 
for both Magellanic Clouds the assumption that all
elements can be  scaled from the solar composition by the same factor is
incorrect. Trundle et al. (\cite{tru07}; hereafter Paper VI) have extended this
sample of narrow lined stars  to $\sim$100 objects and, utilising the same
methods, have derived temperature scales for Galactic and Magellanic Cloud
stars, which imply that stars at lower metallicity have higher effective
temperatures for a given spectral type, in broad agreement with previous 
studies (Martins et al. \cite{mar02, mar05}; Crowther et
al. \cite{cro02,cro06}; Massey et al. \cite{mas04,mas05}). Hunter et al.
(\cite{hun08b}, hereafter Paper VII) have presented chemical compositions for
135 B-type stars in the Large Magellanic Cloud (LMC) with  a broad range of
rotational velocities (up to $\sim$350\,km\,s$^{-1}$). This was the first
significant abundance analysis of rapidly rotating early B-type stars  and they
utilised the nitrogen abundances to test the theory of rotational mixing,
finding the theory to be unable to explain several aspects of the sample. In
particular they found populations of unenriched fast rotators, highly enriched
slow rotators and supergiants that are highly enriched compared to normal
core-hydrogen burning objects (see Sect.~\ref{s_discuss}).  In this paper we
utilise identical methodologies to Paper VII and extend the chemical composition
analysis to the Galactic and Small Magellanic Cloud (SMC) samples from the
FLAMES  large survey and compare these to the LMC stars. 

In Sect.~\ref{s_obs} we briefly describe the survey and
the selection criteria for the objects that are analysed here. In 
Sect.~\ref{s_analysis} a summary of the analysis methodology is presented and 
the chemical compositions of the objects in each metallicity
regime are compared and discussed. In
Sect.~\ref{s_discuss} the nitrogen abundances of the sample are used to test the validity of the conclusions
made in Paper~VII and to further constrain the theory of rotational mixing. 
Finally in Sect.~\ref{s_conclude} we present our principle findings and lay out the challenges for future
theoretical models.

\section{Observations}                                             \label{s_obs}

The observations of the Galactic and Magellanic Cloud stars from the VLT-FLAMES
survey of massive stars have been described in detail in Paper~I and Paper~II 
respectively. To summarise, the majority of the data were obtained using the
Fibre Large Array Multi-Element Spectrograph (FLAMES) on the 8.2\,m European
Southern Observatory Very Large Telescope (ESO-VLT) at Paranal, Chile. These
data were supplemented by UVES and FEROS observations of the brighter
targets. Approximately 300 Galactic objects were observed
and these are associated with the clusters
NGC\,6611, NGC\,3293 and NGC\,4755. Due to the distance of the Magellanic 
Clouds and constraints on fibre placement with the FLAMES instrument, it was
not possible to solely observe cluster objects
and hence the observed Magellanic Cloud samples are dominated by the field populations (see Paper
IV). These field samples were centred towards N\,11 and NGC\,2004 in
the LMC and NGC\,346 and NGC\,330 in the SMC, with over 400 objects being 
observed. The signal-to-noise (S/N) ratios of the SMC spectra were generally in the range 25-150 while
the other spectra were in the range 50-200\footnote{The reduced spectra are publicly available at http://star.pst.qub.ac.uk/$\sim$sjs/flames/}.

In Fig.~\ref{f_spec} (only available online) we provide the observed spectra 
for a subset of the LMC stars.
The strongest nitrogen line (at 3995\AA) has been highlighted and the H$\alpha$
profile is also shown where available.
Panels (a) to (d), (e) to (h) and (i) to (l) 
show objects with low projected rotational velocities (less
than 100\,km\,s$^{-1}$), high projected rotational velocities (greater than
200\,km\,s$^{-1}$) and supergiants respectively. In each of these groups two
relatively nitrogen normal and two nitrogen rich stars are displayed. 
Inspection of the
spectra reveals that apart from the nitrogen lines 
there is no apparent difference between the spectra of
nitrogen rich and nitrogen normal stars.

\addtocounter{figure}{1}
\subsection{Selection criteria} \label{s_select}

Paper~IV presents atmospheric parameters and projected rotational velocities
for the SMC sample of stars from the FLAMES survey.  We have examined the
B-type stars in this  sample of objects and estimated chemical compositions
where possible. Following  the methodology adopted for the LMC sample, upper
limits to the nitrogen abundances have been  estimated when no  nitrogen lines
were observed for the SMC objects. Stars were excluded from the analysis if
asymmetries in the line were obvious or if the upper limits to the
equivalent width estimates for the nitrogen lines lead to abundance estimates
that did not provide a useful constraint. Note that due to spectral
contamination issues Paper~IV does not provide atmospheric parameters for the
double-lined  spectroscopic systems or the Be-type stars and hence no attempt
has been made to derive abundances for these objects here.  Additionally, due
to limitations in the model atmosphere grid used to derive the abundances, the
majority of the O-type stars (i.e. stars hotter than 35\,000\,K) have not been
analysed (see  Sect.~\ref{s_analysis}). These selection criteria are
consistent with those  for the LMC sample (Paper~VII). We note that the
two B8 supergiants analysed in Paper~VI, NGC2004-005  and NGC2004-007, were
excluded as their oxygen abundances may be unreliable.

\begin{figure}[th]
\centering
\epsfig{file=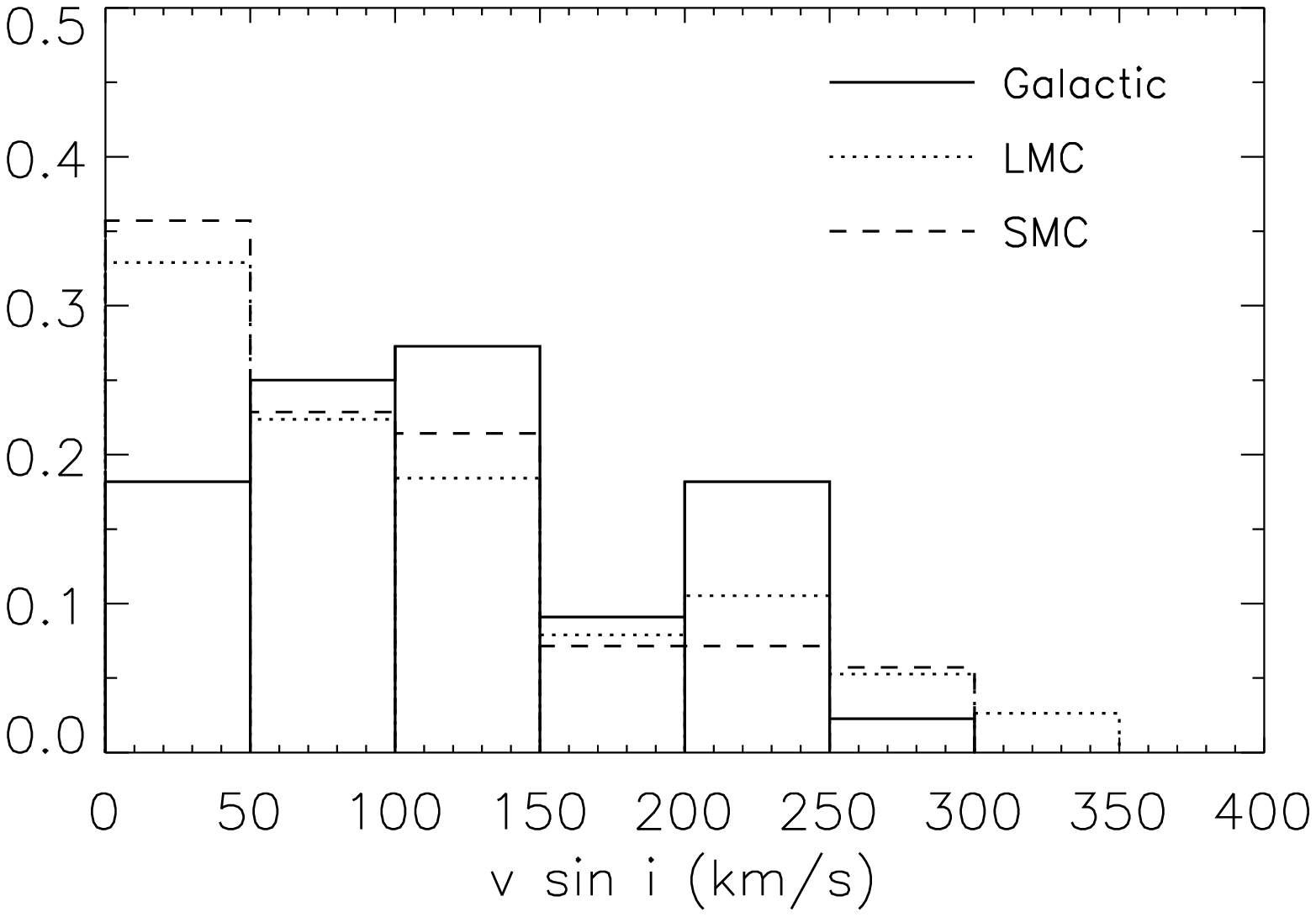, height=60mm, angle=0}
\caption[]{Histogram of the projected rotational velocity distribution of the Galactic,
LMC and SMC samples, normalised to the number of stars in the sample. The three
samples have similar distributions save for a slight deficit of slow rotators in the Galactic sample.}
\label{f_vsinidist}
\end{figure}

In addition to the Magellanic Cloud samples, we require a Galactic comparison
sample. However, given that we are primarily interested in using nitrogen
to constrain the possible enrichment process (see Sect.~\ref{s_intro}),
chemical compositions are only presented for
those Galactic stars with atmospheric parameters in Paper~III and measurable
nitrogen features. Although this introduces a bias towards slow rotators in the Galactic sample, the
upper limits to the nitrogen abundances for the faster rotators were too high to be useful when
comparing with evolutionary models. For the Galactic sample nitrogen lines were observable
at projected rotational velocities up to $\sim$250\,km\,s$^{-1}$.
In Fig.~\ref{f_vsinidist} the projected
rotational velocity distributions of the selected stars in the three regions are plotted. 
Despite the selection criteria, the three distributions are similar although we observe
a smaller proportion of slowly rotating Galactic objects. 
This may be due to Galactic cluster stars rotating faster
than field stars (see Strom et al. \cite{str05}; Paper~III; Huang \& Gies \cite{hua06a}; 
Wolff et al. \cite{wol07}). Hence our Galactic and Magellanic Cloud samples can be considered to be
comparable, with our selection criteria compensating for the intrinsic differences between the Galactic
cluster and the Magellanic Cloud field star velocity distributions.

Papers~V and VI have presented the abundances for the majority of the slow rotators in these 
samples and these abundances have been adopted here. Those stars which have previously
been identified as radial velocity variables (Papers~I, II, III and IV) are considered to
be binaries in the subsequent discussion.

\subsection{Equivalent width measurements}

As discussed in Papers V and VI the equivalent widths (EWs) of the absorption lines 
for the narrow lined stars were measured by fitting Gaussian profiles to the observed
features. The uncertainties in these measurements are typically estimated to be of the order of 10\% 
for well defined lines (Paper~V). However, for faster rotating stars (typically those with projected rotational velocities greater than 50\,km\,s$^{-1}$) the rotational broadening dominates over the
other broadening mechanisms and the use of a Gaussian profile is no longer satisfactory. Hence we 
have utilised rotationally broadened profiles to estimate the equivalent widths of the faster rotators in
our sample. 

The measurement of the metal line equivalent widths for the fast rotators is not as straightforward as the measurements for the slowly rotating stars analysed in Papers~V and VI and many of the associated
problems and uncertainties are discussed in Vrancken et al. (\cite{vra97}).  For example, at high rotational
velocities part of the line can be 'hidden' in the continuum. This obviously will affect the
estimation of the continuum level, the line strength and the line width and can bias the measurements
towards lower equivalent widths. However, a priori knowledge of the projected rotational velocity can
help to alleviate these problems. The rotational velocity of the fast rotators have been
estimated from the helium lines (Papers III and IV) where these problems are
less important due to the strength of these lines. In the estimation of the equivalent widths of the metal lines
we have forced the fits to the lines to have a width equivalent to the projected rotational velocity, and
hence the line width is not underestimated. Additionally, in defining the continuum level the 
continuum is set beyond the region of the broad lines and is fitted with a low order polynomial.

\begin{figure*}
\centering
\epsfig{file=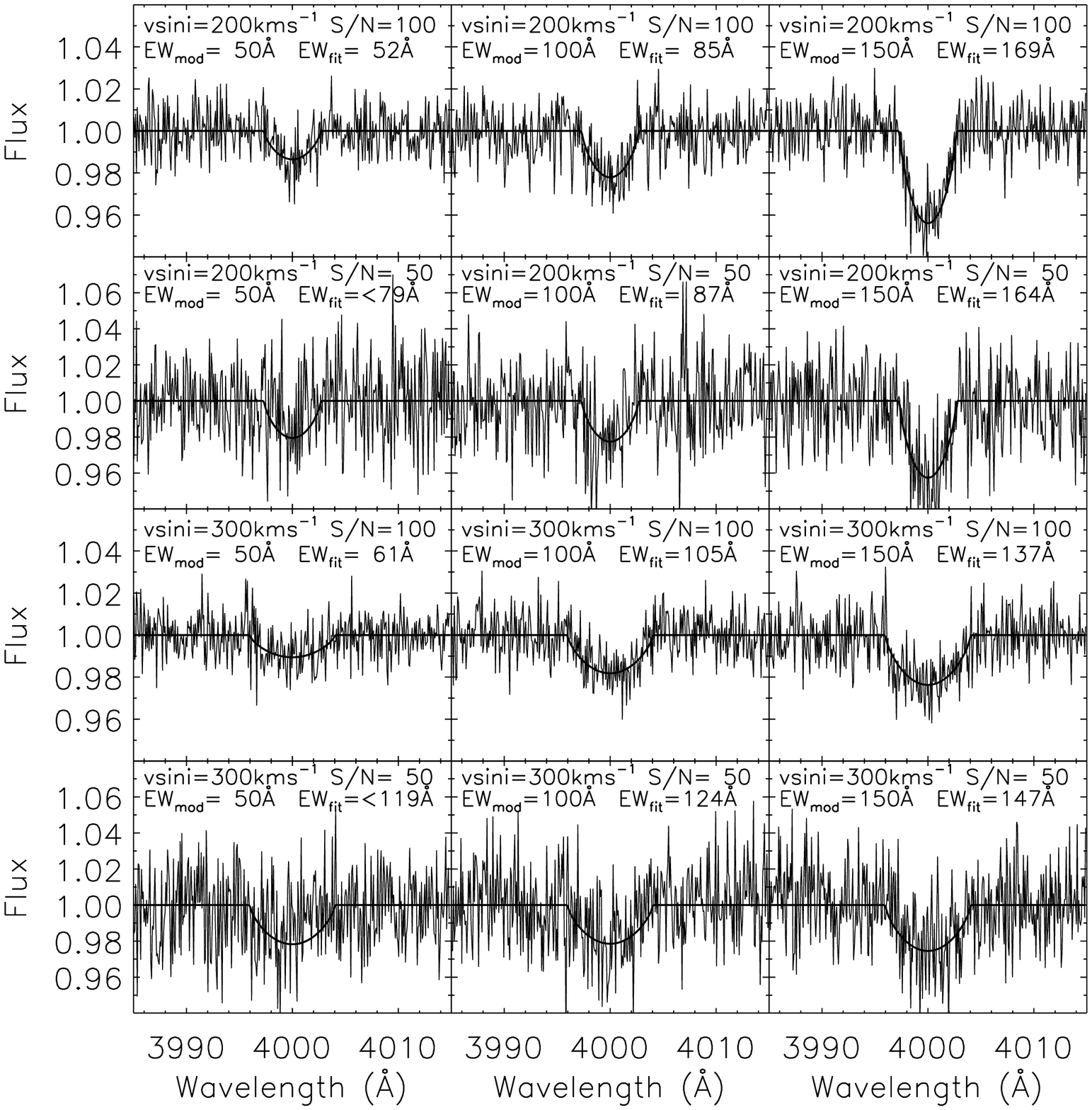, height=180mm, angle=0}
\caption[]{Examples of the quality of the equivalent width measurements for fast rotators at differing S/N ratios. The theoretical spectra
of an absorption line (with an equivalent width of EW$_{\rm mod}$)
is rotationally broadened and noise added to the spectra. Using our IDL fitting routine the equivalent width of this simulated line is then measured (EW$_{\rm fit}$). It should be noted that at a S/N ratio of 50, an absorption line of 50\,m\AA\ could not be observed at these rotational velocities and therefore upper limits to the equivalent width are given. }
\label{f_ew}
\end{figure*}

In Fig.~\ref{f_ew} examples are shown to demonstrate the accuracy of the equivalent width measurements in fast rotating stars for lines of different strength and spectra of differing quality. A theoretical line profile of a known equivalent width has been rotationally broadened and random noise is added to the spectra. This line is then fitted with a rotationally broadened profile. The centroid and width of the line are known quantities and hence these parameters can be constrained. We can therefore estimate the uncertainty in the equivalent width measurements, excluding continuum fitting, to be of the order of $\sim$20\%. Although we have carefully defined the continuum region to be beyond the region of the line, we estimate, by performing repeated fits to a sub-sample of spectra, that the normalisation procedure contributes an additional error of $\sim$20\% to the equivalent width measurements. Hence for the faster rotators in our sample the equivalent width errors can be considered to be of the order of 20-40\%.

The blending of lines in rotationally broadened spectra also affects 
the equivalent width measurements. We again define the
continuum beyond the range of the blended lines and fit the blended lines
simultaneously using multiple rotationally broadened profiles. By constraining
the fitted lines to have the appropriate centroids, rotational broadening
errors in the fitting procedure can be minimised. If this procedure does not
result in a visually satisfactory fit the lines have not been included for
further analysis. However, weak absorption features which are not included will
obviously affect the measured equivalent widths especially for the faster
rotators and would bias the measurements to higher equivalent widths. Comparison of the
abundances between the slow rotators (where blending is not an issue) with
the faster rotators reveals no trend with rotational velocity. Hence we can
conclude that while weak absorption lines may bias our equivalent width 
measurements, such a bias is negligible
within the other uncertainties in our analysis.

The absorption lines used in this analysis are summarized in Paper~V with only
measurable C, O, Mg and Si lines being included. However, given that we are 
primarily interested in the nitrogen abundances, even upper limits to the nitrogen
abundance of a star can be useful for constraining the chemical mixing processes
in the Magellanic Cloud data sets. Hence, as discussed in Sect.~\ref{s_select} 
for the Magellanic Cloud samples, upper limits to the equivalent width of the 
\ion{N}{ii} 3995\AA\ line (which is the strongest nitrogen feature in our spectra)
and to the corresponding nitrogen abundance were estimated. Examples of the fitting of upper limits are also shown in Fig.~\ref{f_ew}. The equivalent widths of the lines included in this analysis along with the estimated abundances (see Sect.~\ref{s_abunds}) are given in Table~1 (only available online).

\section{Analysis}                                            \label{s_analysis}

\subsection{Atmospheric parameters and rotational velocities}

The atmospheric parameters and rotational velocities
have been estimated 
in Papers~III and IV and for completeness a brief overview of
the method is given here. The non-LTE {\sc tlusty} model atmosphere grid 
(see Hubeny \& Lanz \cite{hub95} and references therein) has been used
and is described in detail in 
Dufton et al. (\cite{duf05})\footnote{See also http://star.pst.qub.ac.uk}. This grid covers the effective
temperature range 12\,000--35\,000\,K and allows for the analysis of all the
B-type stars observed in the VLT-FLAMES survey, both the core-hydrogen burning and
supergiant objects. 

As described in Papers~III and IV the hydrogen and helium lines were utilised to
estimate the effective temperatures and surface gravities of the majority of the
sample. For those stars where the helium lines were not temperature sensitive,
effective temperatures were estimated based on spectral type using the effective
temperature calibrations from Paper~VI. The hydrogen
lines were used to constrain the surface gravities in all cases.

For the narrow-lined spectra, where two ionization stages of silicon could be
observed, the effective temperature was determined from the silicon ionization
balance. The majority of these stars have been analysed in Papers~V and VI. For
those stars in which two ionization stages were available, but which lay outside
the selection criteria of these papers, the effective temperatures (and if
necessary the surface gravities) have been re-determined, rather than using
those estimated in Papers~III and Paper~IV. However these parameters are in good agreement
with those previously determined.

Rotational velocities have generally been estimated from the profile fitting
methodology, where a theoretical spectrum is rotationally broadened to fit the
observed spectrum. Again these values are given in Papers~III and IV. We note that for
several cases it was found that the rotational velocity estimated from the helium
lines (Papers~III and IV) did not well fit the metal lines and in these few cases the projected
rotational velocity has been re-determined from the metal lines and is preferred. For the
supergiants in the sample, where macroturbulent broadening can dominate over
rotational broadening, the Fourier Transform technique has been utilised. This
technique has the advantage that rotational broadening can be
separated from other broadening mechanisms (see Sim\'{o}n-D\'{i}az  
\& Herrero \cite{sim07} and Paper~IV).

\subsection{Abundance determinations} \label{s_abunds}

Carbon, nitrogen, oxygen, magnesium and silicon abundances
have been estimated using the atmospheric parameters and EW measurements. In
order to constrain the microturbulence, it has been assumed that the silicon
abundance should be invariant across each metallicity regime and the
microturbulent velocity has been fixed to achieve this where possible. A detailed
description of this methodology is given in Paper~V.
The chemical composition of each star is given in Table~\ref{t_results} (only
available online). This table lists the stellar identifier (from
Papers I and II), atmospheric parameters (effective temperature, surface
gravity and microturbulence), projected rotational velocity 
and chemical composition (C, N, O, Mg and Si)
along with the number of absorption lines observed for each species.

Abundance uncertainties have been estimated using a similar method to that described in 
Paper~V. Random uncertainties have been estimated from the scatter in the line-by-line
abundances of each element and hence include both uncertainties in the EW
measurements and random errors in the atomic data. For those objects where only one line
was observed, the random uncertainties were assumed to be equivalent to the
scatter in the oxygen abundances. If only a few oxygen features were observed
(and
hence it was not possible to estimate this scatter), a random uncertainty of
0.2\,dex was assumed, which is typical for stars with many observed oxygen lines. 
Systematic errors in the abundances were estimated by
considering the uncertainties in each of the atmospheric parameters. For stars
where the effective temperature was constrained from the silicon ionization
balance an uncertainty of 1\,000\,K was adopted, otherwise an uncertainty of
1\,500\,K was assumed. However in contrast to Paper~V we consider that the effective
temperature and surface gravity estimates are correlated in order to
improve our error estimates, with an increase of 1\,000\,K 
resulting in a higher gravity estimate of 0.1\,dex. In addition we adopt measurement
errors in the surface gravity to be 0.05\,dex for stars with gravities of less
than 3.0\,dex and 0.1\,dex otherwise. 
Similarly the microturbulence was better
estimated at small values and an uncertainty of 3\,km\,s$^{-1}$ has been adopted
for values of less than 10\,km\,s$^{-1}$ with 5\,km\,s$^{-1}$ being taken
otherwise. The systematic and random uncertainties were summed in quadrature
to give the uncertainties quoted in Table~\ref{t_results}.

The derived carbon abundances are known to be susceptible to non-LTE
effects (see Sigut \cite{sig96}; Nieva \& Przybilla \cite{nie06}, \cite{nie07})
and these are not fully taken into account by the rather
simplistic model ion that was included in our {\sc tlusty} calculations.
Paper~V discussed adding a correction of 0.34\,dex (from Lennon et al.
\cite{len03}) to the derived carbon abundances from the 4267\AA\ line. This was to achieve
better agreement between the estimated carbon abundance of
B-type stars and those of \ion{H}{ii} regions. However, since the non-LTE
effects will be temperature dependent, such a correction will also be
temperature dependent. Here we adopt a more sophisticated method.

\begin{figure}[th]
\centering
\epsfig{file=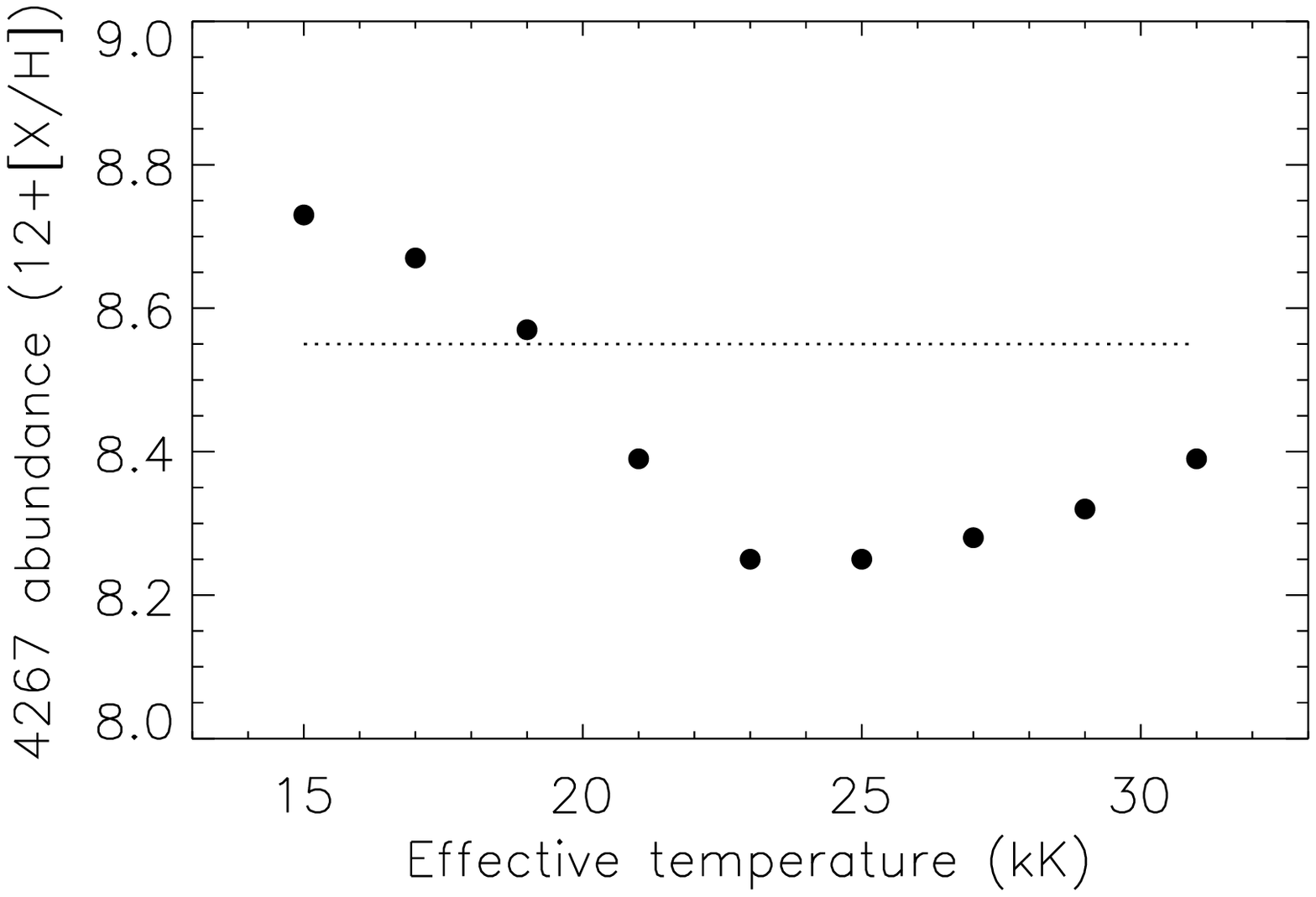, height=60mm, angle=0}
\caption[]{Comparison between the abundance estimated from the \ion{C}{ii} 4267\AA\
line using the {\sc tlusty} grid (points) and that adopted by 
Sigut (\cite{sig96}; for a gravity of 4.00\,dex, microturbulence of 5\,km\,s$^{-1}$ at solar 
metallicity; dashed
line). Corrections have been applied to remove the temperature dependence of the
{\sc tlusty} carbon abundances.
}
\label{f_sigut}
\end{figure}

Sigut (\cite{sig96}) utilises a more complex model atom than does our {\sc TLUSTY} grid.
We have compared abundances from our grid with those calculated
by Sigut over a range of effective temperatures by adopting the equivalent widths that
Sigut used to reproduce a carbon abundance of 8.55\,dex (Grevesse et al. \cite{gre94}). 
The results are shown in Fig.~\ref{f_sigut}. It is clear that there are differences
of up to 0.3\,dex and we have applied corrections to the abundances
from the 4267\AA\ line in order to reproduce the temperature
independent trend of Sigut. Our adopted carbon abundances are solely from
these corrected estimates for this line. We note that the calculations of Sigut
apply only at a surface gravity of 4.00\,dex, a microturbulence of 5\,km\,s$^{-1}$ and
solar metallicity and hence the extrapolation to other parameters and metallicity regimes
should be treated with caution.

\subsection{Chemical compositions}                                \label{s_chem}

In each metallicity regime, the sample of stars is composed of objects in at
least two fields or clusters (see Sect.~\ref{s_obs}). Papers~V and VI have
presented a  detailed abundance analysis of the narrow-lined targets in each
field and show that, within the same metallicity regime, there are no systematic
differences between the abundances from each field. This is consistent with each
Magellanic Cloud being homogeneously mixed and with the Galactic clusters having
similar galactocentric distances (the Galaxy is known to have a  metallicity
gradient, see, for example, Rolleston et al. \cite{rol00}).  In this context we
combine the targets in each metallicity regime in order to increase the sample
size.  In Table~\ref{t_avabunds} the average abundances for each metallicity
regime are given. These are compared to the baseline chemical compositions  from
\ion{H}{ii} regions and the solar composition as recently  determined by
Asplund et al. (\cite{asp05}).

\setcounter{table}{2}
\begin{table*}
\centering
\caption[]{Mean abundances (excluding upper limits) for each metallicity  region
given on the scale 12+$\log$[X/H]$^{1}$. Solar and \ion{H}{ii} region abundances are given for comparison. }
\label{t_avabunds}
\begin{tabular}{lccccccc}
\hline \hline
Species & Solar$^2$ & \multicolumn{2}{c}{Galaxy} & \multicolumn{2}{c}{LMC} &
                                                      \multicolumn{2}{c}{SMC} \\
        &       & B-stars & \ion{H}{ii} regions$^3$ & B-stars & \ion{H}{ii} regions$^4$&
	                                        B-stars &  \ion{H}{ii} regions$^4$\\
\hline \\
Carbon$^5$     & 8.39&8.00$\pm$0.19 (56)&--  &7.70$\pm$0.19 (132)&7.81$\pm$0.22&7.30$\pm$0.28 (81)&7.20$\pm$0.02\\
Nitrogen$^6$   & 7.78&7.62$\pm$0.12 (45)&7.57$\pm$0.30&7.13$\pm$0.29 (~83)&6.92$\pm$0.14&7.24$\pm$0.31 (36)&6.56$\pm$0.07\\
Oxygen     & 8.66&8.63$\pm$0.16 (54)&8.70$\pm$0.30&8.34$\pm$0.13 (133)&8.37$\pm$0.09&7.99$\pm$0.21 (80)&8.02$\pm$0.04\\
Magnesium  & 7.53&7.25$\pm$0.17 (52)&    &7.05$\pm$0.14 (134)&    &6.72$\pm$0.18 (72)&    \\
Silicon    & 7.51&7.42$\pm$0.07 (56)&    &7.17$\pm$0.10 (135)&    &6.77$\pm$0.09 (88)&    \\
\\
\hline
\end{tabular}
\begin{itemize}
\item [$^{1}$] {Values in brackets indicate the  number of
stars for which abundances were estimated.}
\item [$^{2}$] {Asplund et al. \cite{asp05}}.
\item [$^{3}$] {Average of solar neighbourhood regions from Shaver et al. (\cite{sha83})}
\item [$^{4}$] {Kurt \& Dufour (\cite{kur98}}
\item [$^{5}$] {The carbon abundances have been corrected, see text.}
\item [$^{6}$]  {Supergiants have been excluded from the mean nitrogen abundances.} 
\end{itemize}
\end{table*}

\begin{figure*}[ht]
\centering
\epsfig{file=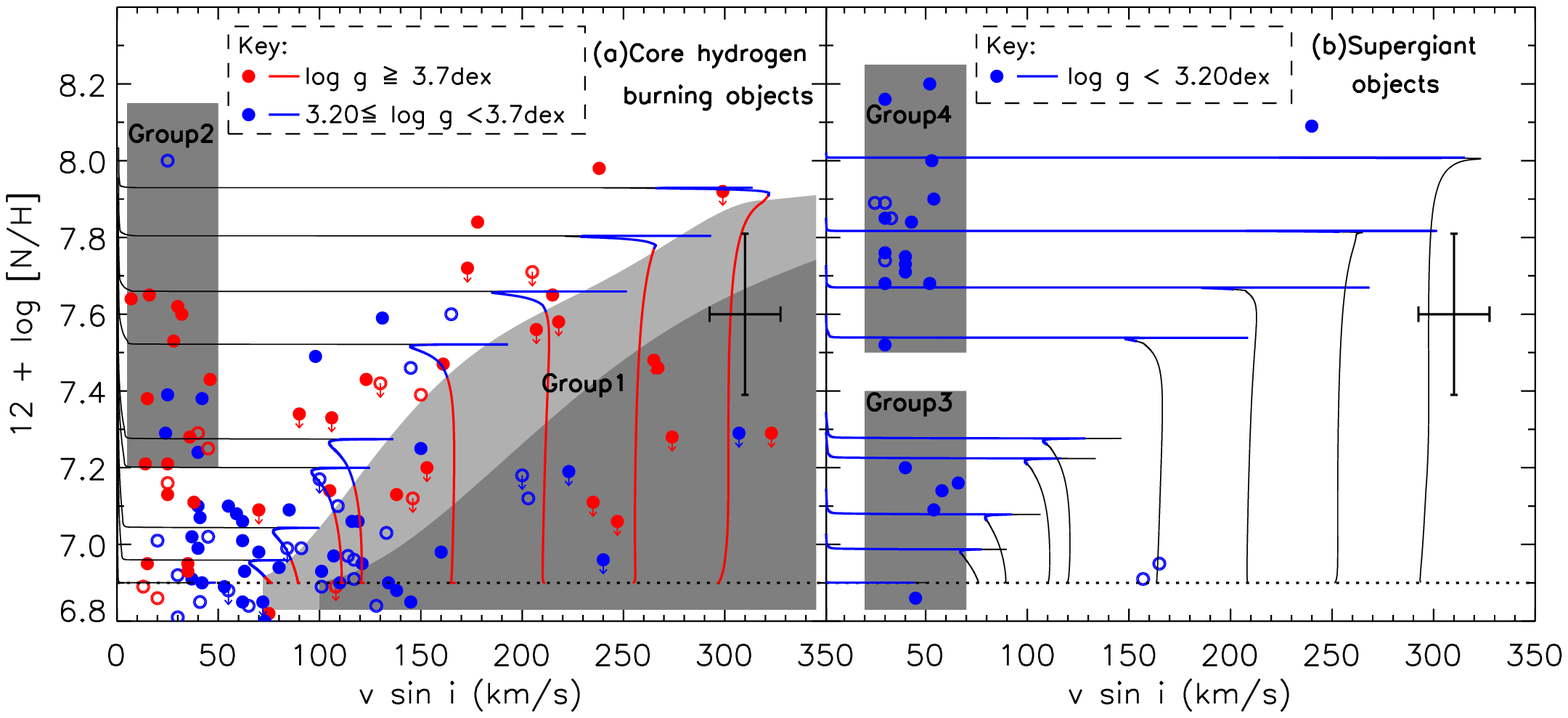, height=80mm, angle=0}
\caption[]{Nitrogen abundance (12+$\log$[N/H]) as a function of projected
rotational velocity for the LMC sample of stars. This plot has been taken from
Paper~VII. The mean error
bars of 
the nitrogen abundances and projected rotational velocities are indicated.
Core-hydrogen burning objects (a) and supergiants (b)
are defined
based on their surface gravity. In panel (a) the core-hydrogen burning objects
are sub-divided to indicate objects in the first part or latter part of the core
hydrogen burning phase, red and blue symbols respectively, approximately 70\% and 30\%
of the main-sequence lifetime.
Open symbols represent radial velocity
variables. Downward arrows indicate upper limits to the nitrogen abundances. The
dotted line indicates the LMC baseline nitrogen abundance as determined from
\ion{H}{ii} region analyses and unmixed B-type stars (Paper~V). The
rotational velocity of the evolutionary models have been scaled by $\pi/ 4$ 
to take into account random angles of inclination. The individual
groups of objects have been discussed in Paper~VII and are summarized in the
text although we have redefined the Group~1 stars (see text). The 
evolutionary models in (a) and (b) have initial masses of 13\,M$_{\rm \sun}$ and
19\,M$_{\rm \sun}$ respectively. We note that a minor error in the temperature 
scale adopted in Paper~VII was discovered and hence several of the points in 
the figure have been corrected. However this is an effectively negligible 
correction as the mean nitrogen abundance and the number of stars in each 
group remain unchanged.}
\label{f_LMC}
\end{figure*}

The mean oxygen abundances of the three regions are in excellent
agreement with the \ion{H}{ii} region abundances. Our Galactic
nitrogen abundance is also in good agreement with that from
\ion{H}{ii} regions. However, the mean nitrogen 
abundance of the SMC from B-type stars is clearly larger than that
from \ion{H}{ii} regions. The LMC nitrogen abundance from B-type stars is also greater than 
that from \ion{H}{ii} regions and, although these mean LMC values can be reconciled within the
uncertainties, it is clear from Sect.~\ref{s_discuss} that there is a mix of normal and significantly enriched stars.
In addition, the scatter
in the stellar nitrogen abundance is larger than that for other elements and
we conclude that this is not dominated by observational scatter,
but due to significant abundance
differences amongst the sample stars (see Sect.~\ref{s_discuss}). 

Our Magellanic Cloud carbon abundances are in good agreement
with those from \ion{H}{ii} regions implying 
that our correction to the carbon abundance may be reasonable. However,
obtaining a Galactic \ion{H}{ii} region carbon
abundance from the literature is highly problematic. For example, 
Dufour et al. (\cite{duf82}) give a value of 8.46\,dex for the
Orion nebula whereas Walter et al. 
(\cite{wal92}) find a range of 7.94--8.78\,dex for the same region
depending on the consideration
of temperature fluctuations. Given the additional problem of determining
accurate stellar carbon abundances it is difficult to confirm the 
validity of our mean Galactic carbon abundances. 

\section{Results}                                           \label{s_discuss}

Chemical compositions are presented for 53 Galactic and 96 SMC
stars and compared with the results for the 135 LMC stars from Paper~VII. In order
to investigate the role of rotational mixing, a large population of fast rotators
is necessary. Our targets have projected rotational velocities up to 
$\sim$300\,km\,s$^{-1}$ and hence are well suited to such a comparison. In particular the nitrogen
abundance can be used as a tracer of surface enrichments and our discussion 
will only consider this element. While in principle associated carbon depletions
would be expected, these would be of the order of the uncertainties in our carbon abundance estimates. 
We adopt the baseline nitrogen abundances of the three regions from Paper~V, which are based
on \ion{H}{ii} regions and unenriched B-type stars. In Paper~VII the
distribution of the nitrogen abundances of the LMC stars 
as a function of their projected rotational velocity 
was described. In Fig.~\ref{f_LMC} the plot from Paper~VII is shown and the 
principle conclusions are summarized below.

\begin{itemize}

\item The majority of the core-hydrogen burning stars show little nitrogen
enrichment and the efficiency of rotational mixing in the evolutionary models
has been constrained to fit the mean nitrogen abundance of this population.  
Many of the nitrogen abundances for the faster rotators are upper limits. We find
that the rotational mixing models provide a reasonable fit to half of our 
core-hydrogen burning data-set, but is in conflict with the rest of the stars (Groups~1 and 2).\\

\item  Group~1 consists of
non-enriched fast rotators, of which many 
are in the late stage of core-hydrogen burning (with relatively
low gravity) and hence a significant nitrogen enrichment
would be expected. This group has been redefined since Paper~VII. The observed sample
was magnitude selected thereby excluding low mass stars close to the beginning of their
core-hydrogen burning phase. We therefore should be biased against young unenriched stars, indicated by the dark shaded area
of Group~1 (plotted for the mean mass of the sample, 13\,M$_{\sun}$). In addition to these stars, the observed evolved stars (blue points) in the light grey region are unexpected as these stars 
are predicted to have higher nitrogen abundances for their rotational velocities and evolutionary
state. Group~1 comprises $\sim$30\% of the non-binary sample, 
two thirds of which are close to the end
of their core-hydrogen burning lifetime (blue points).\\

\item The large number of significantly enriched stars in Group~2 (a further
20\% of the sample) cannot be explained as fast rotators seen at low inclination angles if these 
are randomly distributed. 
The models do not predict strong rotational mixing for slowly rotating stars and in
Paper~VII we postulate that these may be magnetic
stars, equivalent to those discussed by Morel 
et al. (\cite{mor06}, \cite{mor07}). We note that we are not referring to magnetic fields produced 
by dynamo action (Spruit \cite{spr02}) which would be expected to be present in all stars, but rather
fields of fossil origin (see Morel et al. \cite{mor07} and references therein). Huang \& Gies (\cite{hua06b})
also suggest that remnant magnetic fields can account for the peculiar helium abundances
of a subset of their sample. \\

\item The supergiant objects appear to be split into two distinct groups.
The stars in Group~3 (see Fig.~\ref{f_LMC}) have enrichments consistent with the core-hydrogen
burning  phase --- which might be expected if they are their direct descendants. 
However, the supergiants shown in Group~4 have highly enriched
atmospheres. It is postulated that these stars
have previously gone through a red supergiant phase.
As discussed in Paper~VII, our stellar evolution models do predict sufficient enrichment 
in the red supergiant phase, but the location of these Group~4 objects
on the Hertzsprung-Russell diagram cannot
be reproduced.
\end{itemize}

The above summary of the results presented in Paper~VII assumes that 
selection effects in our sample do not introduce a significant observational
bias. One such source of
bias could be that the rapidly rotating stars will appear brighter when viewed pole on (with
low projected rotational velocities) than at high inclination angles (with large
projected rotational velocities). This is due to the pole caps being hotter than
the equatorial regions as discussed by von Zeipel (\cite{vonZ24}). Such a bias
(see also Maeder \cite{mae09})
would affect the Magellanic Cloud samples (which are magnitude limited) and lead to an
oversampling of rapidly rotating pole-on stars (which would populate Group~2
discussed above) and an under-sampling of stars with large projected rotational
velocities that could undergo significant nitrogen enhancement and be the
precursors to the Group~4 supergiants.

We have investigated this possibility as follows. Martayan et al. 
(\cite{mar06,mar07}) estimated mean equatorial velocities for B-type stars
(i.e. excluding the Be-type stars as has been done for the current sample)
in the SMC and LMC ranging from approximately 120 to 160\,km\,s$^{-1}$ 
with similar values being found in Paper\,IV. For Galactic targets, the 
estimated mean values are similar
(Strom et al. \cite{str05}, Huang and Gies \cite{hua06a}, Paper\,III).
For the observed B-type stellar populations discussed by
Martayan et al. only 8\% of their targets have projected rotational
velocities of greater than 300 \,km\,s$^{-1}$, in both NGC\,330 and 
NGC\,2004. Additionally only two stars (out of a sample of more than 300
objects) were found with a projected rotational velocity of greater 
than 400 \,km\,s$^{-1}$. Similar results were found for the 
Magellanic Cloud samples considered here. We have therefore adopted 
a rotational velocity of 400\,km\,s$^{-1}$ as representative of the most
rapidly rotating B-type stars and hence appropriate for investigating the 
possible consequences of the von Zeipel effect for our sample. We note that 
this velocity is consistent or larger than the values that have been assumed when 
modelling rotational mixing in B-type stars (see, for example, 
Heger \& Langer \cite{heg00a}; Meynet \& Maeder  \cite{mey00}).

Adopting typical physical parameters for an early-B-type main
sequence star ($T_{\rm eff}$\,=\,25000\,K; $M = 13$\,M$_{\odot}$;
$R{\mbox{(pole)}} = 5\,$R$_{\odot}$) yields a ratio of the equatorial
rotational velocity to the breakup velocity of $v_{\rm eq}/v{\mbox{(crit)}} =
0.7$ for $v_{\rm eq} = 400$~km/s. For this ratio we estimate the
change in $B$ magnitude between pole-on ($i=0\degr$) and equator-on
($i=90\degr$) views to be only 0.36\,mag, using the methodology outlined
by Townsend et al.\ (\cite{tow04}). This $\Delta{B}$ estimate has
very little dependence on B spectral subtype, and hence should be
representative of our entire B-type main-sequence sample.

The above example implies that there could be a bias in our sample in the sense
that we would preferentially observe fast rotators at low inclination angles.
However it should be noted that such objects will be intrinsically very rare for
two reasons. Firstly the results of Martayan et al. (\cite{mar06}) and
Paper IV indicate that rapidly rotating LMC B-type stars will be 
intrinsically rare,
with those having rotational velocities greater than or equal to 
400\,km\,s$^{-1}$ making up less than 1\% of an unbiased sample of
LMC targets. Additionally the Group 2 stars in 
Fig. \ref{f_LMC} have projected rotational velocities of less
than 50\,km\,s$^{-1}$. Assuming that they have a unique rotational velocity of  
400\,km\,s$^{-1}$, the probability of observing them at such a low angle of
inclination would be less than 1\%; for higher rotational velocities
the probability would be even smaller. Hence 
we estimate that for an LMC population of early B-type stars,
the fraction of stars with a rotational velocity greater than or equal to 
400\,km\,s$^{-1}$ but observed to have a projected rotational velocity of less
than 50\,km\,s$^{-1}$ would be of the order of one in ten thousand. It was the
rareness of such objects that formed the basis of the assumption in Paper~VII
that most of the Group~2 stars that make up one in five of our sample
must be slowly rotating.

We can estimate the amount of bias in our Group~2 sample based on the simulation
discussed above. Our approach has been to exclude all stars from the Group~2
sample that lie within 0.36 magnitudes of our photometric cutoff. Hence if  all
Group 2 stars were rotating at 400\,km\,s$^{-1}$, there would be no bias with
respect  to similar objects observed equatorially ($i= 90\degr$). Exclusion of
such stars reduces the LMC Group 2 sample from seventeen to thirteen stars. 
Given our sample size, we
would expect on average approximately 0.01 Group~2 stars in our sample compared
with the 13 targets found in our corrected sample. Another approach is just to
assume  that our Group~2  stars are fast rotators and that the von Zeipel effect
is far larger than we  have calculated. Then we would expect there to be a
corresponding very large  number of stars (of the order of 10$^{5}$) with large
projected rotational  velocities below our photometric cut-off. The LMC
investigation of Martayan  et al. (\cite{mar06}) went approximately two
magnitudes fainter than our  survey but neither in their photometry nor in their
projected rotational velocity estimates is there any evidence for these objects.

We note that in our simulations, we have made a number of arbitrary assumptions,
for example a single rotational velocity. However relaxing this assumption would
be unlikely to change our conclusions. For example, although there would be more
stars with lower rotational velocities, the von Zeipel effect would be smaller;
conversely more rapid rotators would be very rare as would be the
probability of observing them at low angles of inclination. In conclusion, we
find that even at rotational velocities as high as 400\,km\,s$^{-1}$, the pole
on rotators will appear brighter in the B band by less than 0.4 magnitudes,
whilst only approximately one in ten thousand LMC B-type stars should have 
such rotational velocities with a projected rotational velocity consistent 
with our Group~2 sample. On the basis of these simulations, we conclude that 
almost all, if not all, LMC Group~2 stars must be intrinsically slow rotators.

\subsection{Evolutionary models}                                \label{s_models}

As described in Paper~VII stellar evolutionary models 
have been generated to fit the observed LMC data. Briefly, these models 
have been based on the code of Yoon et al. (\cite{yoo06}) and include the 
effects of rotation (Heger et al. \cite{heg00}) and angular momentum transport
via magnetic torques (Spruit \cite{spr02}). The code has been updated to include
the mass-loss rates of Vink et al. (\cite{vin01}). The magnetically induced
chemical diffusion term is not considered as it is not observationally supported
(Spruit \cite{spr06}). An appropriate chemical composition has been chosen, in
particular with the light elements set to the observed values from Paper~V. The
overshooting parameter (0.335 of the pressure scale height) has been adopted to
reproduce the observed trend of rotational velocities from Paper~IV. The
efficiency parameter for rotational mixing has been set as described in Paper~VII.
In summary, LMC models were calibrated to best fit the observed data set. 

Further evolutionary models have been generated to represent Galactic and SMC
compositions with the overshooting and rotational mixing efficiency parameters
unchanged as these parameters are not predicted to be metallicity dependent. The
light element composition appropriate for these metallicities has again been
adopted from Paper~V. The full grid of evolutionary models will be discussed in
a forthcoming paper (Brott et al. in prep). Here we select models with masses
appropriate for our early B-type stellar sample and which cover a range of
rotational velocities suitable to make comparisons between the current theory
and our observational data set.

\begin{figure*}[ht]
\centering
\epsfig{file=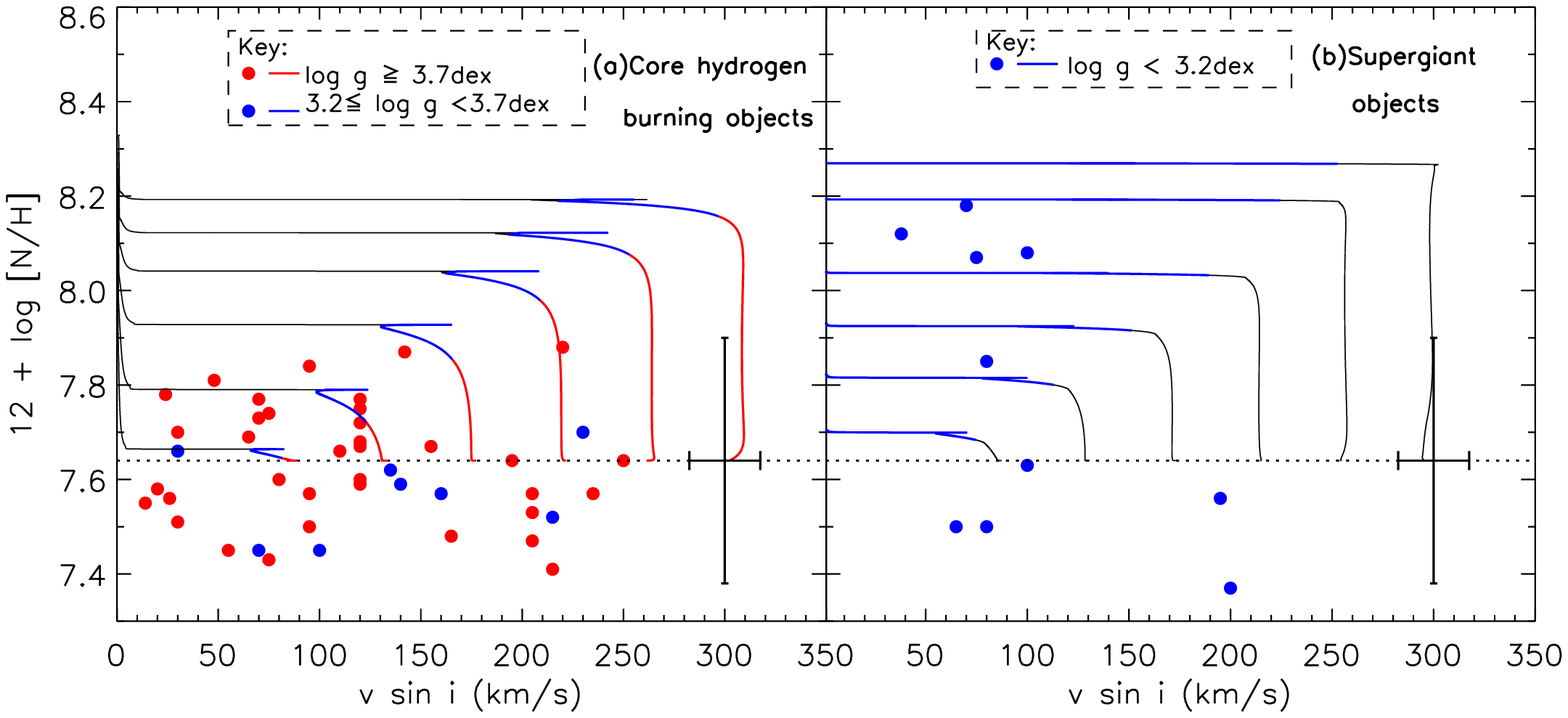, height=80mm, angle=0}
\caption[]{Nitrogen abundance (12+$\log$[N/H]) as a function of projected
rotational velocity for the Galactic sample of stars. Symbols are 
equivalent to those in Fig.~\ref{f_LMC}. Evolutionary models are plotted
for 10 and 20\,M$_{\rm \sun}$ in panels (a) and (b) to represent the mean
mass of the two samples.}
\label{f_GAL}
\end{figure*}

\begin{figure*}[ht]
\centering
\begin{tabular}{c}
\epsfig{file=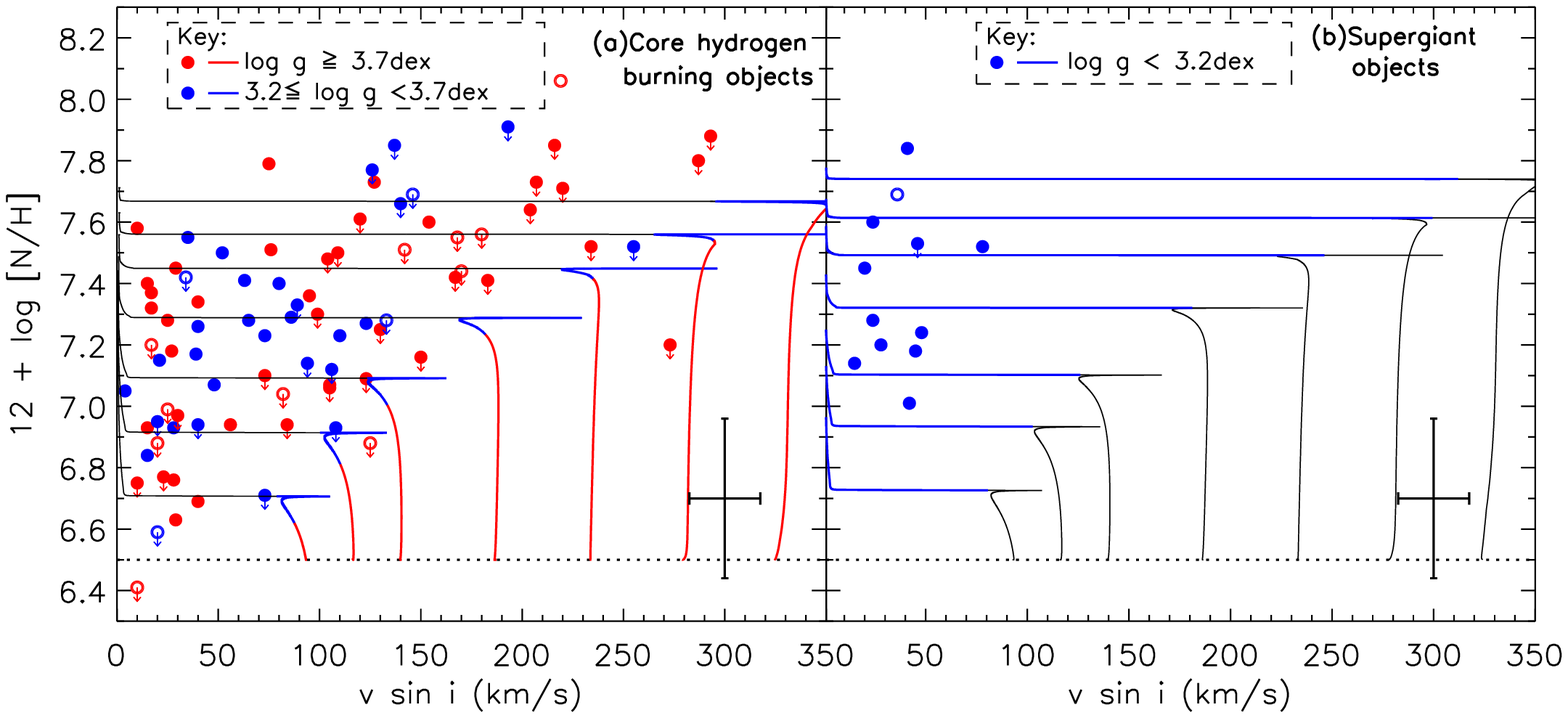, height=80mm, angle=0}\\
\epsfig{file=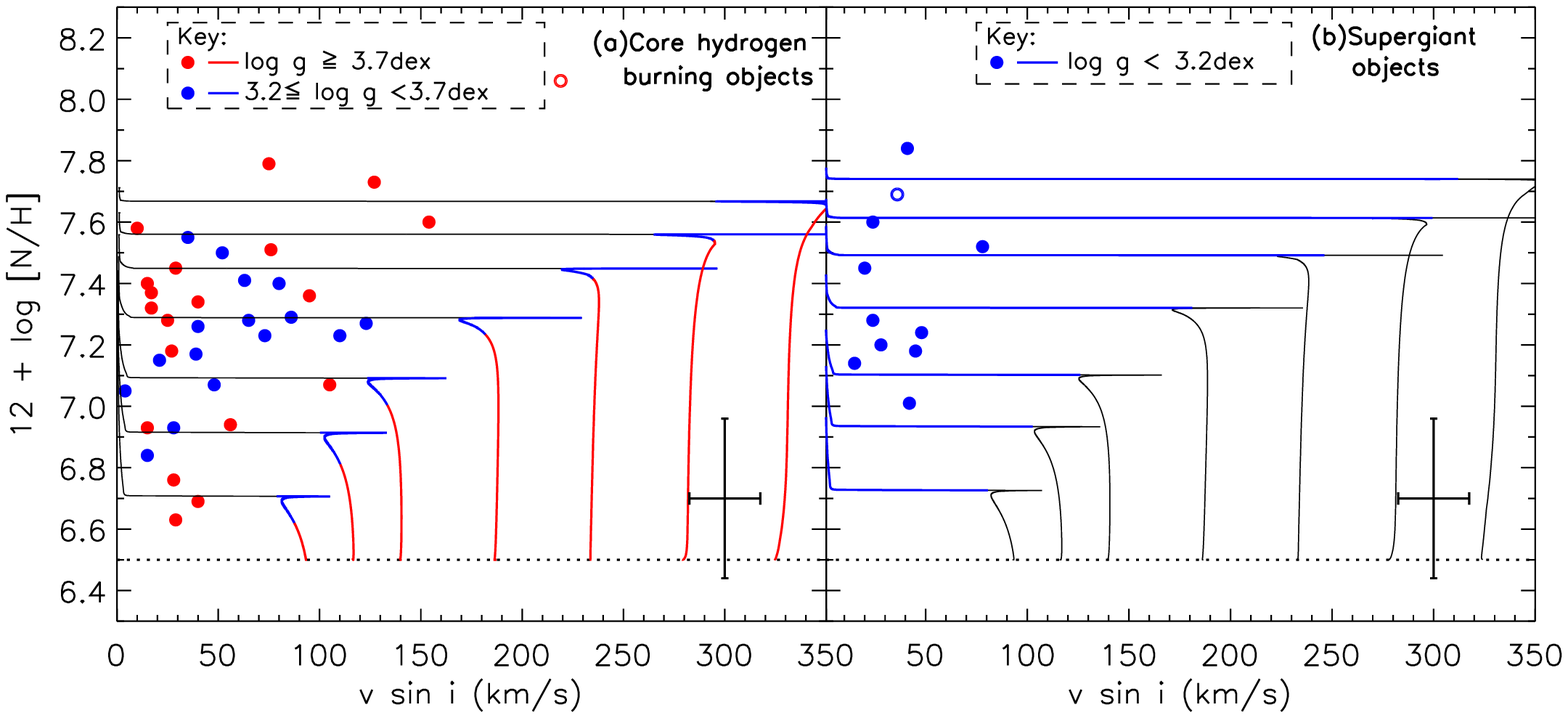, height=80mm, angle=0}\\
\end{tabular}
\caption[]{Nitrogen abundance (12+$\log$[N/H]) as a function of projected
rotational velocity for the SMC sample of stars. 
Symbols are equivalent to those in Fig.~\ref{f_LMC}. The lower panel is
equivalent to the upper panel except that upper limits to the nitrogen
abundances have been removed. Evolutionary models are plotted
for 12 and 13\,M$_{\rm \sun}$ in panels (a) and (b).}
\label{f_SMC}
\end{figure*}

\subsection{Galactic stars}                                      \label{s_gal}

We plot the nitrogen abundances of the Galactic sample as a function of
projected rotational velocity in Fig.~\ref{f_GAL} together with our evolutionary
models. Fig.~\ref{f_GAL}(a) shows that within the abundance uncertainties 
no trend of nitrogen abundance with rotational velocity is observed.
The evolutionary models suggest that for stars with initial rotational velocities of 
250\,km\,s$^{-1}$ nitrogen abundances up to $\sim$8.0\,dex should be observed. 

The lack of any significant enrichment amongst the core-hydrogen burning
stars is surprising. In a population similar to our observed
LMC sample we would expect some nitrogen-rich
fast rotators, as --- even assuming rotational mixing does not occur --- those stars 
could then be interpreted as spun-up mass gainers in close binaries. Also, we
would expect some enriched slow rotators as found in the LMC, as such stars have
been identified in the Galaxy recently by Morel et al. (\cite{mor06}, \cite{mor07}).
Indeed, in the Galactic sample we have only included those stars with
measurable nitrogen lines and while this introduces a bias against the fastest rotating Galactic
stars, it also provides a bias towards the most enriched stars.

The lack of nitrogen-enriched stars may be due to the
majority of the core-hydrogen burning stars in our sample
being relatively unevolved. As rotational mixing is a continuous process, its effect will be
most prominent toward core-hydrogen exhaustion. Also close binary mass transfer
is expected much more frequently toward the end of hydrogen burning, since the
expansion of core-hydrogen burning stars accelerates in the later stages. 
Finally, the process which leads to nitrogen enhancement in the slowly rotating
early B-type stars is as yet unknown. That we do not find such stars
here might imply that the enrichment time scale is of the order of the main
sequence life time. 

The hypothesis that our sample stars are relatively unevolved is supported by the fact that 
in Fig.~\ref{f_GAL}(a), only eight stars have a surface gravity smaller than 3.7\,dex.
Note that the limiting gravity between the two classes is set the same as for the
LMC core-hydrogen burners, but that the Galactic stars are more metal-rich and have
larger radii (and hence lower gravities) 
than LMC stars of the same mass and evolutionary stage. Despite that, the
LMC sample has a majority of stars with $\log g < 3.7$. 
It should be noted that the observations of the Magellanic Cloud samples have been selected
via a magnitude cut (Paper~II) which prohibits the observation of zero-age main-sequence stars
at the lower masses of the sample. For the Galactic sample this magnitude cut
is low enough that the unevolved main sequence stars are not excluded. Additionally
there are no blue supergiants with a mass close to the 
mean sample mass of our core-H burning stars ($\sim 10\,$M$_{\odot}$) in our
Galactic supergiant sample (Fig.~\ref{f_GAL}b). Finally our Galactic sample is mainly composed of 
cluster stars whereas in the Magellanic Cloud we sample more of the field population which may 
contain more evolved stars.

Our sample shows seven Galactic core-hydrogen burning objects with low
surface gravities (i.e. close to the end of the core-hydrogen
burning phase; blue points) 
and nitrogen abundances lower than that predicted for their evolutionary stage.
The two apparently fast rotating supergiants with low nitrogen abundances could also be placed in this
group within the uncertainty in their surface gravities. According to the models we would expect these
objects to be nitrogen rich, in particular the fastest rotators. These
objects may be analogous to the Group~1 stars shown in Fig.~\ref{f_LMC} and 
make up $\sim$20\% of the core-hydrogen burning sample, although if we adopt the
maximum error several objects could be reconciled with the models. 
While this group is smaller than that observed in the LMC sample, 
given the less evolved nature of our Galactic sample this is not unexpected.

In order to determine if we should observe a group of objects similar to Group~2
in Fig.~\ref{f_LMC}(a) we have added the mean absolute enrichment of those 
objects to the Galactic baseline nitrogen abundance (on a logarithmic scale) and find that such a group
of objects would appear to have a normal nitrogen abundance 
within the abundance uncertainties. However, 
Morel et al. (\cite{mor07}) find slowly rotating magnetic
Galactic stars with
nitrogen abundances up to 8.0\,dex. In Paper~VII we considered
these stars to be analogous to the Group~2 stars in the LMC and linked fossil
magnetic fields to chemical mixing, although the mixing mechanism is
unknown. It is unclear why we do not see these stars in our Galactic
sample. The Morel et al. sample contains some bias towards
such stars and this bias is not present in our Galactic data set. Additionally
our Galactic sample is predominantly composed of cluster stars whereas field  
stars dominate the Magellanic Cloud samples. Wolff et al. (\cite{wol07})
have provided evidence that cluster stars rotate significantly faster than
field stars, which is why our Galactic sample contains fewer slower rotators
than our Magellanic Cloud samples (see Fig.~\ref{f_vsinidist}). They relate this to 
differences in the star formation 
process between cluster and field stars.

We also find some evidence for a difference 
between field and cluster stars in our LMC sample. The LMC targets lie towards 
the clusters NGC\,2004 and N\,11. In the former case almost our entire sample 
lies beyond the cluster radius and hence is probably a field star population, whereas 
for N\,11 our stars are taken from across the association and the field. While $\sim$65\% of
our LMC sample of core-hydrogen burning non-binary stars is taken from the
NGC\,2004 targets, 80\% of the Group~2 stars are from the field of NGC\,2004.
This may indicate that the Group~2 stars are more likely to occur in the field population
and hence explain the lack of such stars in our Galactic cluster sample.
However, the significance of the lack of highly mixed slowly rotating core 
hydrogen burning 
stars cannot be truly tested.

Figure~\ref{f_GAL}(b) suggests that there are two groups of supergiants with different degrees
of nitrogen enrichment,
in agreement with the situation in the LMC sample, Fig.~\ref{f_LMC}(b), although the number
of stars in the Galactic sample is limited. Given
that no core-hydrogen burning Galactic objects are observed with nitrogen
abundance consistent with the highly enriched supergiants, it appears unlikely that
these objects have evolved directly from such a phase and hence they
should not be used to constrain the models of core-hydrogen burning stars.
The supergiant sample is further
discussed in Sect.~\ref{s_super}. 

\subsection{SMC stars}                                           \label{s_smc}

In Fig.~\ref{f_SMC} we plot the nitrogen abundance as a function of projected
rotational velocity for the SMC sample of stars. From the upper panel there
appears to be a strong correlation between these two quantities,
as would be expected from the theory of 
rotational mixing. However, for the majority of stars with 
rotational velocities greater than 100\,km\,s$^{-1}$ only upper
limits to the nitrogen abundances have been estimated and such a limit will be
correlated with the projected rotational velocity leading to the apparent trend.
If we remove these upper limits
(lower panel) this correlation disappears. The lack
of absolute nitrogen abundances for these fast rotators prevents the
observation of a group of stars similar to the Group~1 stars of
Fig.~\ref{f_LMC}(a). We note that although the estimation of upper limits to the nitrogen abundance
is not metallicity dependent, our SMC data is of lower quality (Sect.~\ref{s_obs}) 
and hence higher upper limits are derived, particularly for the fastest rotators.

A comparison between Fig.~\ref{f_SMC}(a) and 
Fig.~\ref{f_LMC}(a) shows that while the majority of the LMC
stars have undergone little or no enrichment, many stars in the SMC 
appear to be significantly enriched.
Our mean SMC nitrogen abundance is actually higher than
that of the LMC despite the SMC having a lower baseline abundance. However,
given that nitrogen lines can only be observed for abundances greater than approximately 6.9\,dex
for even a moderately rotating star, it is difficult to test the significance of
this observation and it should be noted that the mean values are consistent within the uncertainties. Nevertheless, we 
find 30 of our 70 non-binary core-hydrogen burning stars (excluding upper limits)
have such high estimates that, even adopting an
average error bar of $\pm$0.26\,dex, they have nitrogen abundances that are inconsistent with the
core-hydrogen burning regime of the stellar evolution
tracks in Fig.~\ref{f_SMC}(a). This is obviously a lower limit as stars with upper limits to their nitrogen abundances could also populate this group if their abundances are indeed higher
than the model predictions.

The mean nitrogen abundance of the SMC sample is 7.24\,dex and to reproduce this
abundance
evolutionary tracks with an initial rotational
velocity of $\sim$200\,km\,s$^{-1}$ are required. The probabilities of a star
rotating at 200\,km\,s$^{-1}$ appearing to have projected rotational velocities
of less than 100 and 50\,km\,s$^{-1}$ are 13\% and 3\% respectively.  We observe 19 and 12
stars with such projected rotational velocities 
but with nitrogen abundances consistent, within their
uncertainties, with a 200\,km\,s$^{-1}$ evolutionary track. It is therefore
statistically very unlikely that these groups are populated by fast rotators
with low angles of inclination. For example, in order to have 12 stars rotating
at 200\,km\,s$^{-1}$ and appearing with rotational velocities of less than
50\,km\,s$^{-1}$ a population of 400 stars would be required with $\sim$260 of
these having projected rotational velocities in the range 150-200\,km\,s$^{-1}$.
Such a population clearly does not exist in the upper panel of Fig.\ref{f_SMC}
and, although Be-type stars are excluded, 
there is no bias against normal B-type stars with these velocities. We can
therefore conclude that the enriched stars with low projected rotational
velocities have intrinsically low rotational velocities. This is
clearly in conflict with the theory of rotational mixing and these stars
may be analogous to the LMC stars designated as Group~2 in Fig.~\ref{f_LMC}.

There appears to be a gap between the Group~2 stars in the LMC
(Fig.~\ref{f_LMC}a) and higher velocity stars. Such a gap is not clear in the
SMC sample. The significance of this gap is unknown, but if it is real it
implies that, whatever the process that is enriching the Group~2 stars, it is efficient at
higher velocities in the SMC sample. Alternatively the sample may be populated
by two mixing processes, the first being that producing the LMC Group~2 stars
while the latter process mixes all the SMC stars to a similar nitrogen
abundance. Indeed, it should
be noted that the majority of the Group~2 LMC stars are high gravity stars and the
same holds true for the analogous sample in the SMC (e.g. projected rotational velocities of less
than 50\,km\,s$^{-1}$ and nitrogen abundances greater than 7.2\,dex), which is also dominated by a
large population of high gravity stars. 

Our sample could again be biased by the von
Zeipel effect (\cite{vonZ24}). To estimate the degree of bias, we have again
considered an object with an equatorial rotational velocity of 400 km\,s$^{-1}$.
As discussed above the Group~2 objects are less clearly defined in the SMC
sample. However adopting the criteria of a logarithmic surface gravity
greater than $\geq$3.2 dex,  projected rotational velocity of 
$\leq$50 km\,s$^{-1}$ and a nitrogen abundance of $\geq$7.2\,dex (corresponding
to a nitrogen enhancement of more than 0.6 dex), we find 11 stars. Removing
those within 0.36 magnitudes of our B-band magnitude cut-off reduces the 
sample size to 8 stars. SMC B-type stars have intrinsically larger 
rotational velocities than their LMC or Galactic counterparts 
(Martayan et al. \cite{mar07}; Paper~IV) and we would expect approximately 3\%
to have values of 400 km\,s$^{-1}$\, or more. Again less than 1\% of them will 
have an inclination angle leading to a projected rotational velocity of 
$\leq$50 km\,s$^{-1}$. Hence we would expect that approximately one in four
thousand SMC B-type stars would have a rotational velocity of  
$\geq$400 km\,s$^{-1}$ and a projected rotational velocity of 
$\leq$50 km\,s$^{-1}$. In turn, for our SMC sample size, we would then expect 
on average 0.02 such stars to lie in the same region as the Group~2 LMC stars. Alternatively, if we assume
that all our Group~2 sample are rapidly rotating, we would expect a
corresponding  population of approximately 30\,000 rapidly rotating stars
with larger inclination angles. Martayan et al. (\cite{mar07}) study of the 
SMC cluster NGC\,330 goes approximately two magnitudes fainter than our sample
and again there is no evidence for this population. Hence we
conclude that at least several (and probably most, or all) of the
nitrogen-enhanced SMC targets with low projected
rotational velocities must be intrinsically slow rotators.

In both Fig.~\ref{f_LMC}(b) and Fig.~\ref{f_GAL}(b) the supergiant sample consisted of
two groups of objects, one mildly enriched (or unenriched) group consistent with the core-hydrogen burning stars and a highly enriched group. This is not observed in the
SMC sample (Fig.~\ref{f_SMC}b). However, given that almost all the stars in the
core-hydrogen burning sample are significantly enriched, in contrast to the
Galactic and LMC samples, this might mask such a distribution. Although the mass
of the SMC supergiants ($\sim$13\,M$_{\rm \sun}$) is lower than that of the
Galaxy and LMC samples ($\sim$20\,M$_{\rm \sun}$) the correlation of nitrogen abundance with mass
is expected to be small (compare evolutionary tracks in Fig.~\ref{f_LMC} a and b) and none
is observed in the LMC sample of stars which cover a wide range of masses.
Nevertheless there are a
number of supergiant objects with nitrogen abundances greater than 7.6\,dex,
which is higher than the nitrogen abundance generally observed during the core 
hydrogen
burning phase, indicating that part of the mixing in these supergiants may occur after the
hydrogen burning phase. 

\subsection{Supergiants}   \label{s_super}

\begin{figure}[h]
\centering
\begin{tabular}{c}
\epsfig{file=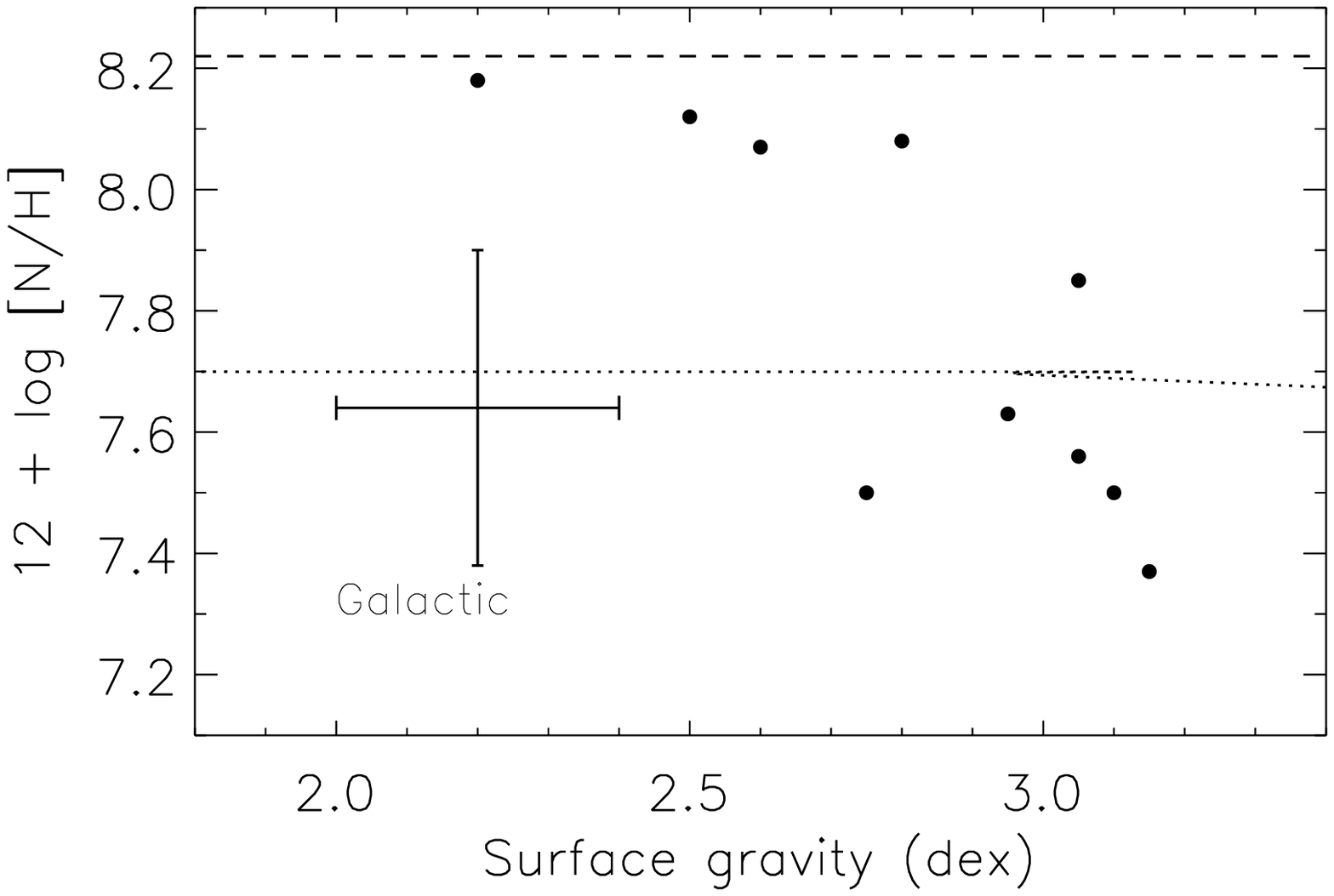, height=55mm, angle=0}\\
\epsfig{file=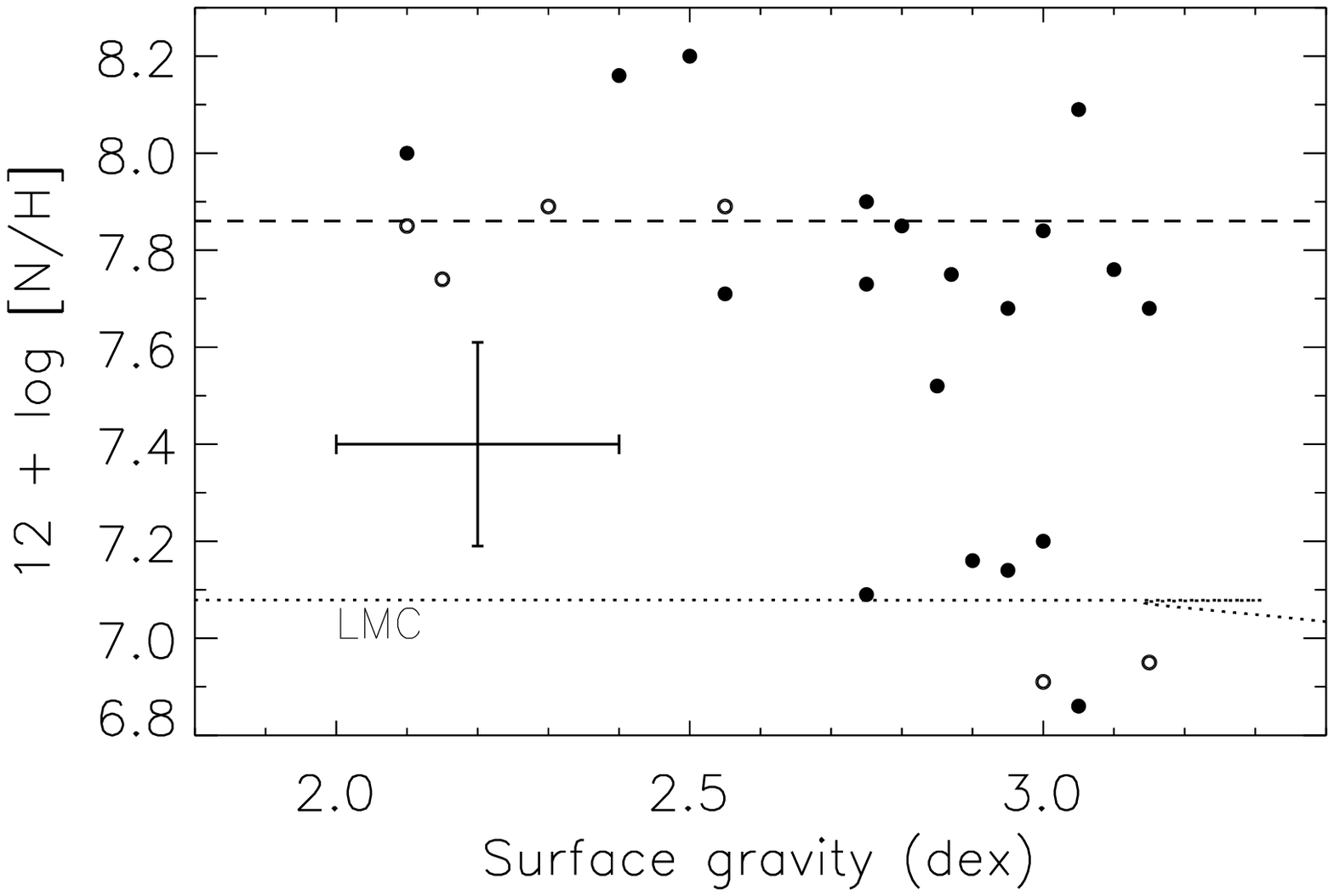, height=55mm, angle=0}\\
\epsfig{file=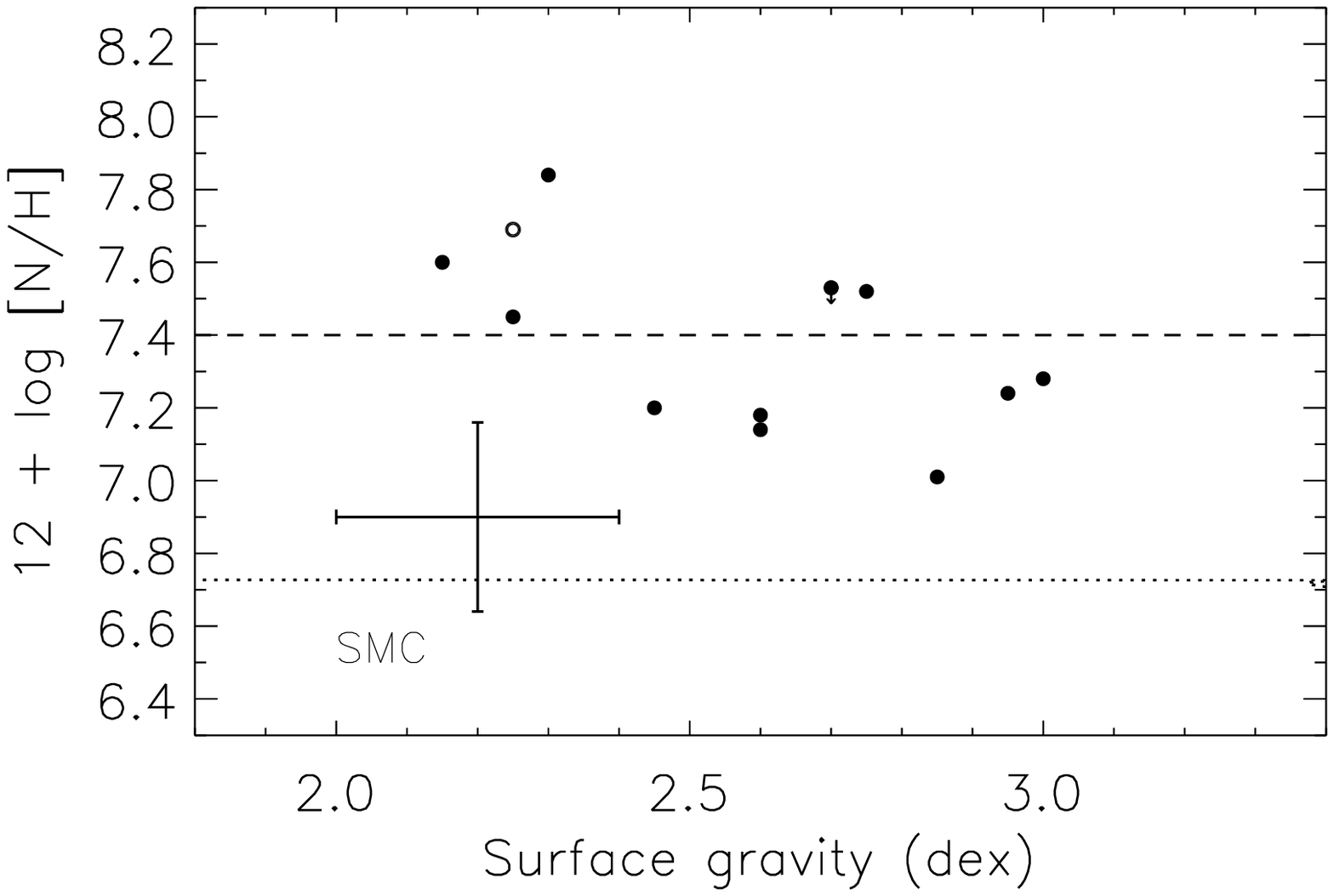, height=55mm, angle=0}\\
\end{tabular}
\caption[]{Nitrogen abundances of the supergiants in each region. Open symbols represent 
radial velocity variables. Error bars represent the mean uncertainty in the nitrogen abundance
for each of the samples and an uncertainty of 0.2\,dex in surface gravity.
The dotted line represents the evolutionary track for a model at
the mean mass of the sample and a moderate rotational velocity (100\,km\,s$^{-1}$) although as the
main-sequence rotational velocity of these stars is unknown it is not possible to make a detailed
comparison with the models. The dashed
line represents the nitrogen abundances achieved during the first dredge-up for the same models,
although it should be noted that the models do not predict a return to the temperatures and
gravities representative for our observations.}
\label{f_supers}
\end{figure}

As discussed above, the supergiants typically show two groups, one highly
enriched group and one less enriched (or normal) group, although the effect is
less obvious in the SMC sample. In Fig.~\ref{f_supers} the nitrogen abundance is
plotted against the surface gravity (evolutionary status) for each of the three
metallicity regimes. It is clear that the lowest gravity supergiants have
the highest nitrogen abundances at all metallicities.

As discussed in Paper~VII and illustrated in Fig.~\ref{f_supers}, 
the amount of surface nitrogen enrichment that 
is obtained during the first dredge-up phase can reasonably reproduce the observed
nitrogen abundances of the enriched supergiants in the LMC. 
This suggests that some of our blue supergiants have previously evolved through
a red supergiant stage, with their surfaces enriched through
dredge-up. Given that there is little predicted trend with mass and velocity (less than
0.3\,dex) for the amount of enrichment during this red supergiant stage, this
would imply that all stars which
have gone through such a phase should have similar nitrogen abundances
regardless of their initial parameters while stars which have not gone through this
phase would have lower abundances. Such bimodal distributions are
clear in the Galactic and LMC samples. The expected nitrogen enrichment from
dredge-up is in excellent agreement with the LMC observations and, while it is
slightly higher than our Galactic observations, within the uncertainties it
is consistent. The models predict that the nitrogen enrichment from dredge-up in the SMC
sample is not significantly greater than that observed for the majority of our
core-hydrogen burning sample and hence a bi-modal distribution would not be
easily observed. The predicted nitrogen abundance after first dredge-up is lower
than that observed for the most enriched SMC supergiants but again, within the
uncertainties, the hypothesis that the most enriched supergiants have gone through a
red supergiant phase is not unreasonable. 

However, in order for our enriched supergiants to be explained by such a process
it would be necessary for those stars to return to hotter temperatures to allow
us to observe them in the blue supergiant stage on a Hertzsprung-Russell diagram. 
Such a blue-loop was, for example, proposed
as the mechanism to explain the progenitor of SN1987A. Whilst these
loops are not predicted by our evolutionary models, those in the Geneva models
(see, for example, Maeder \& Meynet \cite{mae01}) do not extend to high enough
masses or to hot enough temperatures to reproduce our observations. If the
single star nature of these supergiants can be observationally confirmed it is
clear that more extensive blue loops need to be predicted by the evolutionary
models. Indeed, such a scenario may also help to bring the expected and observed
number of blue supergiants into better agreement (Paper~IV). 

Venn \& Przybilla (\cite{ven03}, a re-analysis of previous work, Venn \cite{ven99}) have presented similar nitrogen abundances for A-F type supergiants in the Galaxy and SMC and drawn similar conclusions to those presented above.

\section{Conclusions}                                         \label{s_conclude}

Chemical compositions are presented 
for $\sim$50 Galactic and $\sim$100 SMC
early B-type stars which cover a broad range of rotational velocities and
complement previously published results for $\sim$130 LMC stars of similar
spectral
types for which new observationally constrained evolutionary models had been
generated. We compare our new observational data sets with the LMC results.

\subsection{Core-hydrogen burning objects} 

In both the Magellanic Cloud populations we find a significant excess of
slowly rotating nitrogen enriched stars that cannot be explained within the
context of rotational mixing, at least for a random distribution of inclination 
angles; $\sim$20\% in the LMC and $\sim$40\% in the SMC. In the Galactic sample of the FLAMES survey, such stars
are not found, although they are present in other Galactic samples (Morel et al. \cite{mor06}, \cite{mor07}). 
While this may be attributed to the early evolutionary stage
of the majority of the FLAMES Galactic stars, it would constrain the time scale 
of the enrichment process
for this group --- the physics of which is unknown --- to be comparable
to the main-sequence lifetime. 
 
The LMC sample contains a significant group of stars which are rapidly
rotating, but show no significant nitrogen enhancement. Due to the magnitude cut used
in our target selection, the zero-age main-sequence is not well sampled at low mass. Hence if
only single stars are considered and rotational mixing is assumed to be valid, 
such nitrogen normal rapid rotators are not predicted by rotational mixing models. A significant
fraction of these stars have gravities that imply that they are close to the end of the core-hydrogen
burning phase.
The Galactic sample also reveals several fast rotating low gravity stars although they constitute 
a smaller fraction probably due to the less evolved nature of the Galactic sample.
In the SMC, the rather high upper limits on nitrogen abundances for rapid rotators does not allow
for their identification.

Finally, the LMC sample contains about 20\% of stars which are rather rapidly
rotating and nitrogen enriched. Those stars could either be rotationally mixed
single stars (Maeder et al. \cite{mae09}, Brott et al. \cite{bro09})
if the non-enriched rapid rotators are {\em all} binary products),
or the spun-up mass gainer in mass transfer binaries (Langer et al. \cite{lan08}).
While some such stars may be contained in the SMC sample, they are not found
in our Galactic sample. This may be due to the rather unevolved nature
of this sample although it should be noted that our Galactic sample contains
fewer slow rotators compared to the Magellanic Cloud sample.

The differences in data quality and mean evolutionary state of the samples
do not allow clear trends in relative numbers of the various groups
of core-hydrogen burning stars to be obtained for each metallicity regime. 

\subsection{Supergiants}

In all three metallicity regimes a population of highly enriched 
supergiant objects have been observed, which are unlikely to be the direct
descendants of rotationally
mixed core-hydrogen burning stars. The predicted amount of nitrogen after 
first dredge up reasonably reproduces their nitrogen abundances. 
However, the position of these stars in the Hertzsprung-Russell diagram cannot 
be reproduced by our evolutionary models and, if the single star nature of these objects 
can be confirmed, argues for the occurrence of blue loops at hotter 
temperatures and higher masses than currently predicted.\\

In summary we find that both the Galactic and SMC samples support the
conclusions made in Paper~VII. It should also be noted that we find no
evidence for an excess of binary systems in the groups of objects
that are in conflict with the rotating single star models although our observations do not allow
for complete binary identification.
Our results imply that a new nitrogen enrichment
process (or processes) other than rotational mixing must be important in massive star evolution.
Furthermore, close binary effects may be needed to understand the rapid rotators. 
Whether rotational mixing is required to understand our results remains an open question
at this time, but can be answered by a rigourous study of the binary fraction in
those groups of stars which appear to contradict the theory.
 
\begin{acknowledgements}
We are grateful to staff from the European Southern Observatory 
for assistance in
obtaining the data. 
This work, conducted as part of the award `Understanding 
the lives of
massive stars from birth to supernovae' (SJS) made under the
European Heads of Research Councils and European Science Foundation
EURYI (European Young Investigator) Awards scheme, was supported by
funds from the Participating Organisations of EURYI and the EC Sixth
Framework Programme. SJS also acknowledges the Leverhulme Trust.  Additionally
we acknowledge financial support from
STFC (UK Science and Technology Facilities Council),
NWO (Netherlands Science Foundation) and DEL (Department of Education and Learning
in Northern Ireland).

\end{acknowledgements}


\newpage

\begin{landscape}
\pagestyle{empty}
\addtolength{\topmargin}{60mm} 
\setcounter{figure}{0}
\begin{figure*}
\centering
\begin{tabular}{c}
\epsfig{file=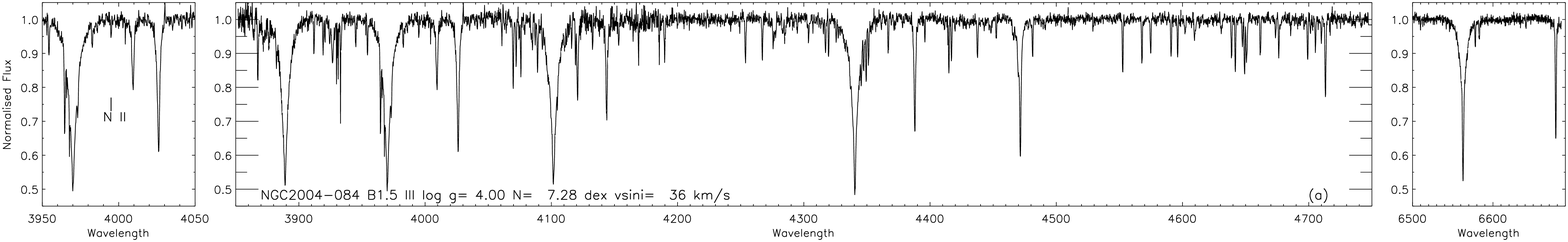, height=40mm, angle=-0} \\
\epsfig{file=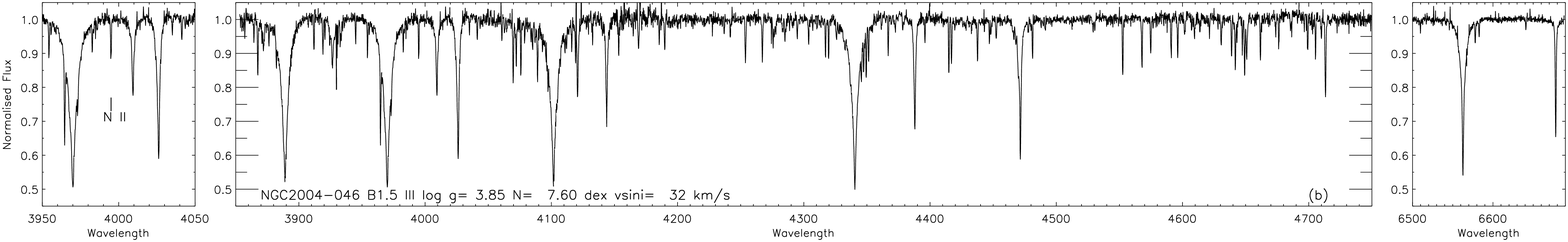, height=40mm, angle=-0} \\
\epsfig{file=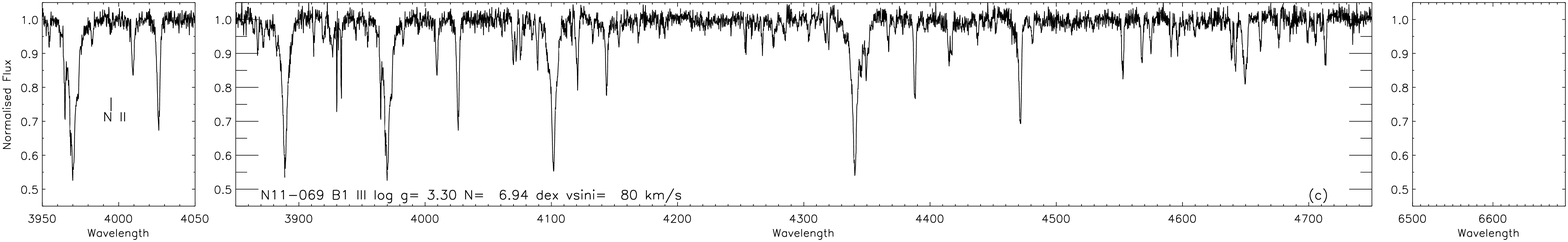, height=40mm, angle=-0} \\
\epsfig{file=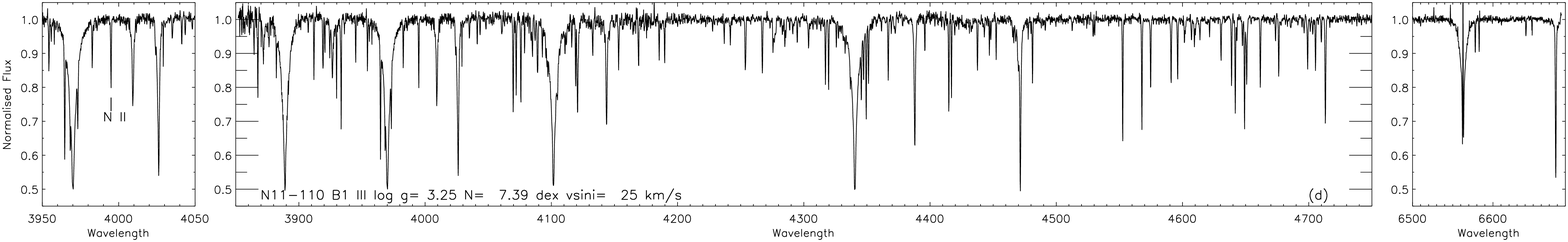, height=40mm, angle=-0} \\
\end{tabular}
\caption[]{Comparison of the spectra for a subset of 
nitrogen rich and nitrogen normal LMC stars. The spectral region containing the
\ion{N}{ii} 3995\AA\ line is expanded (left panel) and the line is 
indicated. The spectral type, surface gravity, nitrogen abundance and 
projected rotational velocity of these stars are indicated in the panels.
Other than the strength of the nitrogen lines there are no systematic 
differences between
the spectra of stars of the same spectral type, surface gravity and projected
rotational velocity. }
\label{f_spec}
\end{figure*}
\end{landscape}

\begin{landscape}
\pagestyle{empty}
\addtolength{\topmargin}{60mm} 
\begin{figure*}
\centering
\begin{tabular}{c}
\epsfig{file=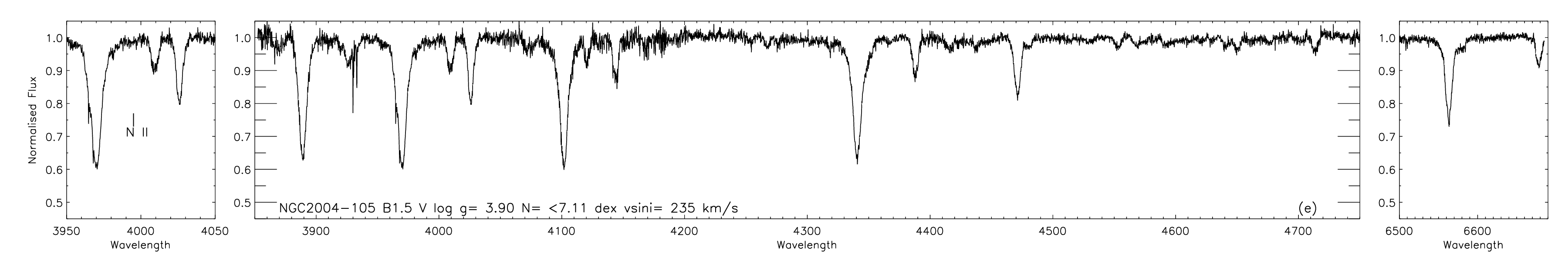, height=40mm, angle=-0} \\
\epsfig{file=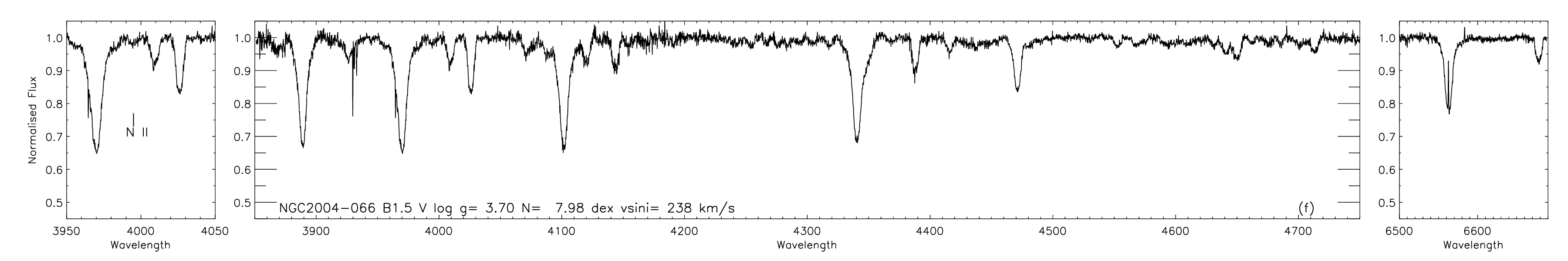, height=40mm, angle=-0} \\
\epsfig{file=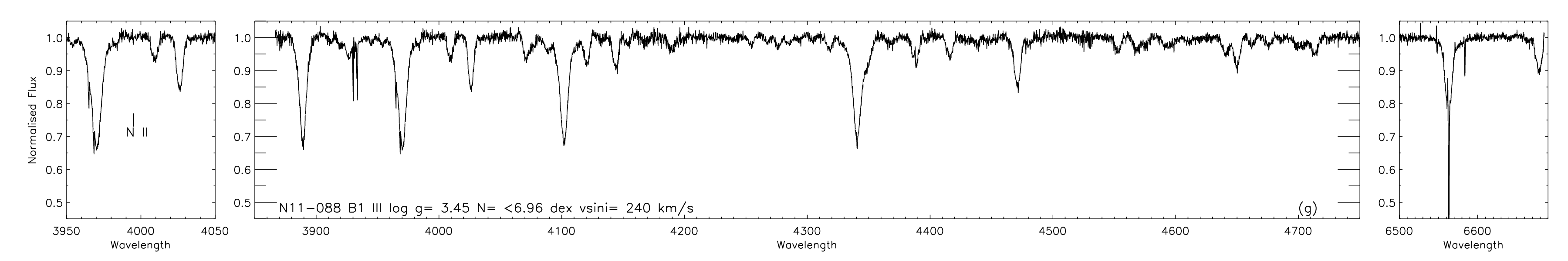, height=40mm, angle=-0} \\
\epsfig{file=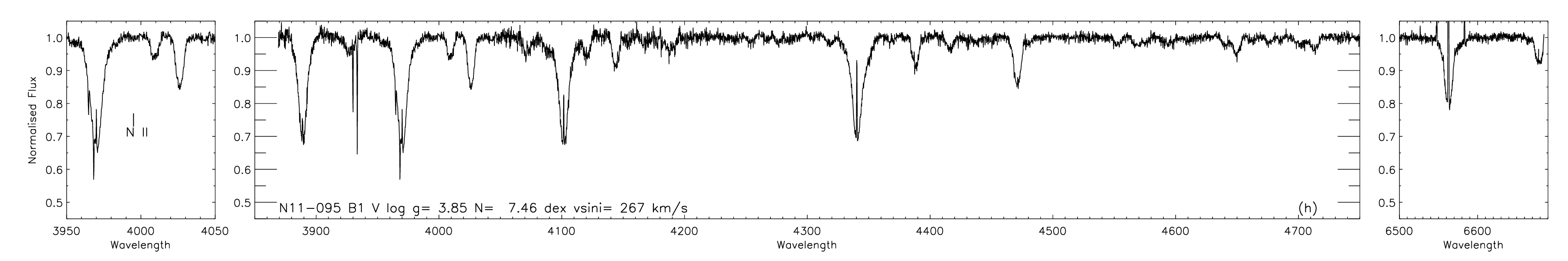, height=40mm, angle=-0} \\
\end{tabular}
\addtocounter{figure}{-1}
\caption[]{-continued.}
\end{figure*}
\end{landscape}

\begin{landscape}
\pagestyle{empty}
\addtolength{\topmargin}{60mm} 
\begin{figure*}
\centering
\begin{tabular}{c}
\epsfig{file=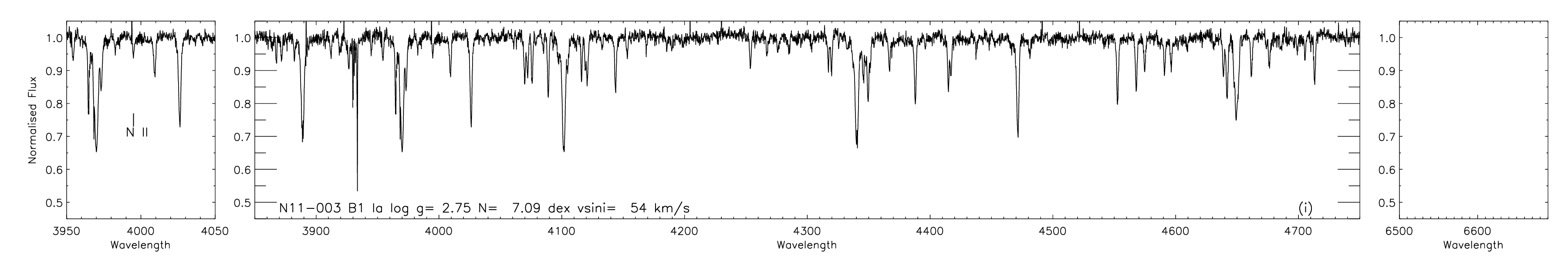, height=40mm, angle=-0} \\
\epsfig{file=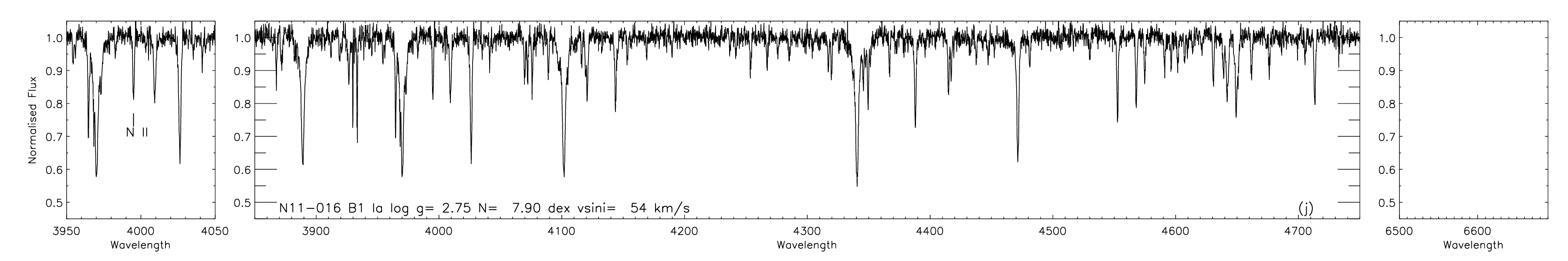, height=40mm, angle=-0} \\
\epsfig{file=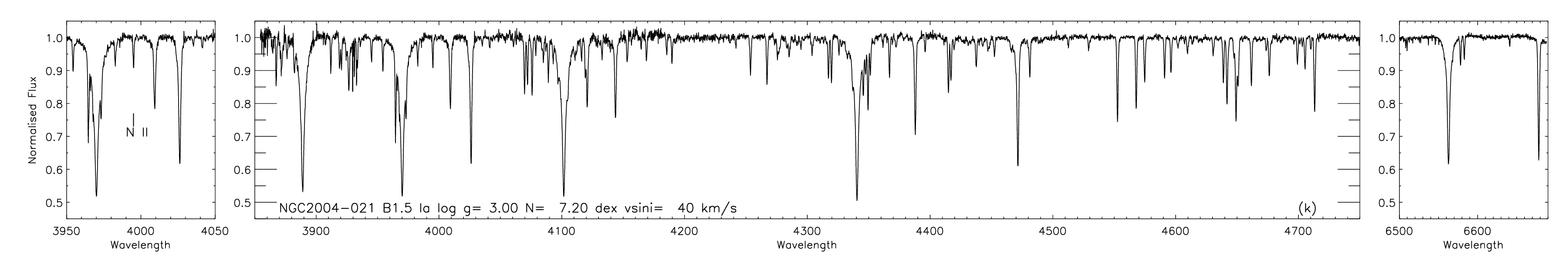, height=40mm, angle=-0} \\
\epsfig{file=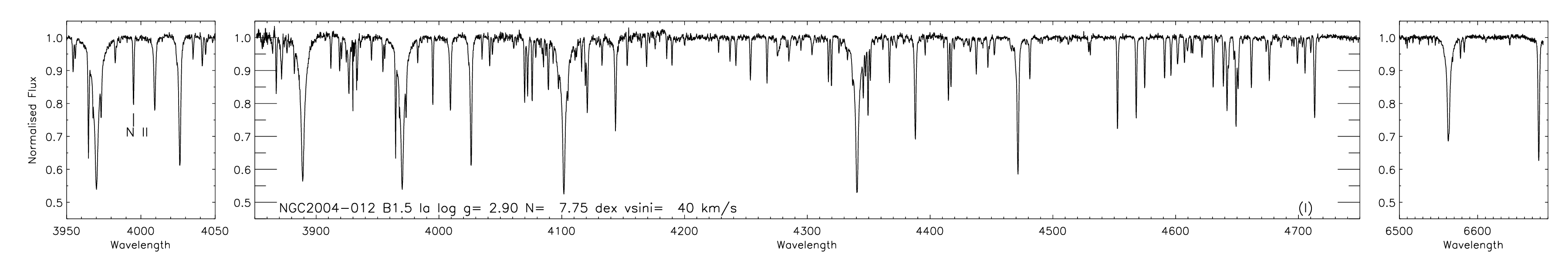, height=40mm, angle=-0} \\
\end{tabular}
\addtocounter{figure}{-1}
\caption[]{-continued.}
\end{figure*}
\end{landscape}

\setcounter{table}{0}
\clearpage
\begin{table*}
\caption[]{Equivalent width measurements and derived abundances for the metal lines observed in the spectra of our sample stars.}
\begin{tabular}{lccccccccccc}
\hline \hline
Species&Wavelength&\multicolumn{2}{c}{NGC6611-001}
&\multicolumn{2}{c}{NGC6611-006}&\multicolumn{2}{c}{NGC6611-012}
&\multicolumn{2}{c}{NGC6611-020}&\multicolumn{2}{c}{NGC6611-021}\\
       &          &EW&Abund&EW&Abund&EW&Abund&EW&Abund&EW&Abund\\
\hline \\
  C II&4267.00 & 88&8.06 & 91&8.14 &134&8.19 &111&7.98 &112&7.92\\
  N II&3955.85 & --&  -- & --&  -- & --&  -- & --&  -- & --&  --\\
  N II&3995.00 & 42&7.52 & 47&7.60 & 74&7.50 & --&  -- & 64&7.43\\
  N II&4447.03 & 48&7.85 & 26&7.54 & --&  -- & 49&7.67 & 35&7.45\\
  N II&4601.48 & 42&8.13 & --&  -- & --&  -- & --&  -- & --&  --\\
\end{tabular}
\end{table*}

\setcounter{table}{0}
\clearpage
\begin{table*}
\caption[]{-continued.}
\begin{tabular}{lccccccccccc}
\hline \hline
Species&Wavelength&\multicolumn{2}{c}{NGC6611-025}
&\multicolumn{2}{c}{NGC6611-030}&\multicolumn{2}{c}{NGC6611-032}
&\multicolumn{2}{c}{NGC6611-033}&\multicolumn{2}{c}{NGC6611-035}\\
       &          &EW&Abund&EW&Abund&EW&Abund&EW&Abund&EW&Abund\\
\hline \\
  C II&4267.00 &157&8.19 &218&8.35 &209&8.32 &160&8.19 &114&8.00\\
  N II&3955.85 & --&  -- & --&  -- & --&  -- & --&  -- & --&  --\\
  N II&3995.00 & 82&7.58 & 69&7.53 & 58&7.39 & 96&7.62 & --&  --\\
  N II&4447.03 & 64&7.85 & 40&7.63 & 40&7.65 & 56&7.65 & 54&7.74\\
  N II&4601.48 & 60&8.09 & 29&7.59 & 43&7.93 & --&  -- & 44&7.90\\
\end{tabular}
\end{table*}

\setcounter{table}{0}
\clearpage
\begin{table*}
\caption[]{-continued.}
\begin{tabular}{lccccccccccc}
\hline \hline
Species&Wavelength&\multicolumn{2}{c}{NGC6611-042}
&\multicolumn{2}{c}{NGC6611-052}&\multicolumn{2}{c}{NGC6611-062}
&\multicolumn{2}{c}{NGC6611-063}&\multicolumn{2}{c}{NGC6611-066}\\
       &          &EW&Abund&EW&Abund&EW&Abund&EW&Abund&EW&Abund\\
\hline \\
  C II&4267.00 &184&8.03 &180&7.74 &181&7.91 &201&8.19 &201&8.18\\
  N II&3955.85 & --&  -- & --&  -- & --&  -- & --&  -- & --&  --\\
  N II&3995.00 & 79&7.43 & 66&7.62 & 36&7.48 & 63&7.33 & 41&7.36\\
  N II&4447.03 & 47&7.57 & --&  -- & --&  -- & --&  -- & --&  --\\
  N II&4601.48 & 61&7.98 & --&  -- & --&  -- & --&  -- & --&  --\\
\end{tabular}
\end{table*}

\setcounter{table}{0}
\clearpage
\begin{table*}
\caption[]{-continued.}
\begin{tabular}{lccccccccccc}
\hline \hline
Species&Wavelength&\multicolumn{2}{c}{NGC3293-001}
&\multicolumn{2}{c}{NGC3293-002}&\multicolumn{2}{c}{NGC3293-003}
&\multicolumn{2}{c}{NGC3293-004}&\multicolumn{2}{c}{NGC3293-005}\\
       &          &EW&Abund&EW&Abund&EW&Abund&EW&Abund&EW&Abund\\
\hline \\
  C II&4267.00 & 98&8.01 &150&8.10 &237&7.95 &214&8.17 &203&7.89\\
  N II&3955.85 & --&  -- & 46&8.10 & --&  -- & --&  -- & --&  --\\
  N II&3995.00 & 97&7.83 &188&8.01 &127&7.41 &115&7.39 &152&7.50\\
  N II&4447.03 & 58&7.84 & --&  -- & 65&7.52 & 78&7.63 & --&  --\\
  N II&4601.48 & 36&7.84 &111&8.22 & 68&7.64 & 69&7.71 & 69&7.62\\
\end{tabular}
\end{table*}

\setcounter{table}{0}
\clearpage
\begin{table*}
\caption[]{-continued.}
\begin{tabular}{lccccccccccc}
\hline \hline
Species&Wavelength&\multicolumn{2}{c}{NGC3293-006}
&\multicolumn{2}{c}{NGC3293-007}&\multicolumn{2}{c}{NGC3293-008}
&\multicolumn{2}{c}{NGC3293-010}&\multicolumn{2}{c}{NGC3293-012}\\
       &          &EW&Abund&EW&Abund&EW&Abund&EW&Abund&EW&Abund\\
\hline \\
  C II&4267.00 &193&7.81 &200&8.13 &195&7.83 &194&7.85 &201&7.86\\
  N II&3955.85 & --&  -- & 27&7.56 & --&  -- & 27&7.51 & --&  --\\
  N II&3995.00 &115&7.29 &123&7.48 &108&7.30 &116&7.37 & 94&7.23\\
  N II&4447.03 & --&  -- & 59&7.49 & 89&7.70 & 49&7.34 & --&  --\\
  N II&4601.48 & --&  -- & --&  -- & 77&7.75 & 39&7.37 & 40&7.40\\
\end{tabular}
\end{table*}

\setcounter{table}{0}
\clearpage
\begin{table*}
\caption[]{-continued.}
\begin{tabular}{lccccccccccc}
\hline \hline
Species&Wavelength&\multicolumn{2}{c}{NGC3293-016}
&\multicolumn{2}{c}{NGC3293-018}&\multicolumn{2}{c}{NGC3293-019}
&\multicolumn{2}{c}{NGC3293-023}&\multicolumn{2}{c}{NGC3293-024}\\
       &          &EW&Abund&EW&Abund&EW&Abund&EW&Abund&EW&Abund\\
\hline \\
  C II&4267.00 &199&8.13 &155&7.96 &180&8.26 &178&7.72 &182&7.75\\
  N II&3955.85 & --&  -- & 22&7.57 & --&  -- & --&  -- & --&  --\\
  N II&3995.00 & 35&7.38 & 94&7.47 & --&  -- & 72&7.30 & 71&7.33\\
  N II&4447.03 & 25&7.81 & 50&7.50 & --&  -- & 56&7.69 & 48&7.63\\
  N II&4601.48 & --&  -- & --&  -- & --&  -- & 40&7.61 & 55&7.90\\
\end{tabular}
\end{table*}

\setcounter{table}{0}
\clearpage
\begin{table*}
\caption[]{-continued.}
\begin{tabular}{lccccccccccc}
\hline \hline
Species&Wavelength&\multicolumn{2}{c}{NGC3293-025}
&\multicolumn{2}{c}{NGC3293-026}&\multicolumn{2}{c}{NGC3293-028}
&\multicolumn{2}{c}{NGC3293-030}&\multicolumn{2}{c}{NGC3293-031}\\
       &          &EW&Abund&EW&Abund&EW&Abund&EW&Abund&EW&Abund\\
\hline \\
  C II&4267.00 &179&7.84 &181&8.03 &178&7.69 &189&7.86 &189&7.66\\
  N II&3955.85 & --&  -- & 26&7.72 & --&  -- & --&  -- & --&  --\\
  N II&3995.00 & --&  -- & 89&7.58 & 50&7.28 & 62&7.51 & --&  --\\
  N II&4447.03 & --&  -- & 55&7.69 & 29&7.54 & 35&7.68 & --&  --\\
  N II&4601.48 & --&  -- & 42&7.67 & --&  -- & 18&7.40 & --&  --\\
\end{tabular}
\end{table*}

\setcounter{table}{0}
\clearpage
\begin{table*}
\caption[]{-continued.}
\begin{tabular}{lccccccccccc}
\hline \hline
Species&Wavelength&\multicolumn{2}{c}{NGC3293-034}
&\multicolumn{2}{c}{NGC3293-035}&\multicolumn{2}{c}{NGC3293-037}
&\multicolumn{2}{c}{NGC3293-038}&\multicolumn{2}{c}{NGC3293-041}\\
       &          &EW&Abund&EW&Abund&EW&Abund&EW&Abund&EW&Abund\\
\hline \\
  C II&4267.00 &157&8.28 &196&7.92 &194&7.99 &211&8.16 &230&8.34\\
  N II&3955.85 & 17&7.66 & --&  -- & --&  -- & --&  -- & --&  --\\
  N II&3995.00 & 64&7.41 & --&  -- & 70&7.62 & 53&7.46 & 40&7.26\\
  N II&4447.03 & 48&7.66 & --&  -- & 35&7.65 & --&  -- & --&  --\\
  N II&4601.48 & 43&7.87 & --&  -- & 45&8.03 & --&  -- & --&  --\\
\end{tabular}
\end{table*}

\setcounter{table}{0}
\clearpage
\begin{table*}
\caption[]{-continued.}
\begin{tabular}{lccccccccccc}
\hline \hline
Species&Wavelength&\multicolumn{2}{c}{NGC3293-043}
&\multicolumn{2}{c}{NGC3293-062}&\multicolumn{2}{c}{NGC4755-002}
&\multicolumn{2}{c}{NGC4755-003}&\multicolumn{2}{c}{NGC4755-004}\\
       &          &EW&Abund&EW&Abund&EW&Abund&EW&Abund&EW&Abund\\
\hline \\
  C II&4267.00 &169&7.94 &162&8.07 &289&7.72 &274&7.70 &240&7.79\\
  N II&3955.85 & 10&7.74 & --&  -- & 66&8.25 & 62&8.05 & 78&8.06\\
  N II&3995.00 & 34&7.35 & 28&7.64 &192&8.06 &219&8.04 &269&7.99\\
  N II&4447.03 & 18&7.56 & --&  -- & 85&8.04 &106&8.01 &140&8.00\\
  N II&4601.48 & 11&7.42 & --&  -- &103&8.24 &118&8.18 &161&8.20\\
\end{tabular}
\end{table*}

\setcounter{table}{0}
\clearpage
\begin{table*}
\caption[]{-continued.}
\begin{tabular}{lccccccccccc}
\hline \hline
Species&Wavelength&\multicolumn{2}{c}{NGC4755-006}
&\multicolumn{2}{c}{NGC4755-015}&\multicolumn{2}{c}{NGC4755-016}
&\multicolumn{2}{c}{NGC4755-017}&\multicolumn{2}{c}{NGC4755-020}\\
       &          &EW&Abund&EW&Abund&EW&Abund&EW&Abund&EW&Abund\\
\hline \\
  C II&4267.00 &223&7.66 &165&7.90 &177&7.98 &203&8.04 &213&8.22\\
  N II&3955.85 & --&  -- & 40&7.96 & --&  -- & --&  -- & 15&7.57\\
  N II&3995.00 &131&7.39 &117&7.74 & --&  -- & 61&7.53 & 67&7.51\\
  N II&4447.03 & --&  -- & 45&7.46 & --&  -- & 52&8.02 & --&  --\\
  N II&4601.48 & 88&7.81 & --&  -- & --&  -- & --&  -- & 40&7.85\\
\end{tabular}
\end{table*}

\setcounter{table}{0}
\clearpage
\begin{table*}
\caption[]{-continued.}
\begin{tabular}{lccccccccccc}
\hline \hline
Species&Wavelength&\multicolumn{2}{c}{NGC4755-023}
&\multicolumn{2}{c}{NGC4755-026}&\multicolumn{2}{c}{NGC4755-029}
&\multicolumn{2}{c}{NGC4755-033}&\multicolumn{2}{c}{NGC4755-040}\\
       &          &EW&Abund&EW&Abund&EW&Abund&EW&Abund&EW&Abund\\
\hline \\
  C II&4267.00 &213&8.17 &212&8.08 &213&8.21 &190&8.07 &219&8.29\\
  N II&3955.85 & --&  -- & --&  -- & --&  -- & --&  -- & --&  --\\
  N II&3995.00 & 40&7.35 & --&  -- & 46&7.47 & 21&7.23 & 39&7.58\\
  N II&4447.03 & 35&7.91 & --&  -- & --&  -- & --&  -- & --&  --\\
  N II&4601.48 & 25&7.82 & --&  -- & 25&7.80 & --&  -- & --&  --\\
\end{tabular}
\end{table*}

\setcounter{table}{0}
\clearpage
\begin{table*}
\caption[]{-continued.}
\begin{tabular}{lccccccccccc}
\hline \hline
Species&Wavelength&\multicolumn{2}{c}{NGC4755-048}
&\multicolumn{2}{c}{NGC346-012}&\multicolumn{2}{c}{NGC346-021}
&\multicolumn{2}{c}{NGC346-026}&\multicolumn{2}{c}{NGC346-027}\\
       &          &EW&Abund&EW&Abund&EW&Abund&EW&Abund&EW&Abund\\
\hline \\
  C II&4267.00 &207&8.22 & 62&7.45 & 63&7.52 & --&  -- & --&  --\\
  N II&3955.85 & --&  -- & --&  -- & --&  -- & --&  -- & --&  --\\
  N II&3995.00 & 24&7.30 & 46&6.94 & 33&6.81 & --&  -- & $<$63&$<$7.71\\
  N II&4447.03 & --&  -- & --&  -- & --&  -- & --&  -- & --&  --\\
  N II&4601.48 & --&  -- & --&  -- & --&  -- & --&  -- & --&  --\\
\end{tabular}
\end{table*}

\setcounter{table}{0}
\clearpage
\begin{table*}
\caption[]{-continued.}
\begin{tabular}{lccccccccccc}
\hline \hline
Species&Wavelength&\multicolumn{2}{c}{NGC346-029}&\multicolumn{2}{c}{NGC346-032}
&\multicolumn{2}{c}{NGC346-037}&\multicolumn{2}{c}{NGC346-039}
&\multicolumn{2}{c}{NGC346-040}\\
       &          &EW&Abund&EW&Abund&EW&Abund&EW&Abund&EW&Abund\\
\hline \\
  C II&4267.00 & 16&7.33 & 42&7.44 & 79&7.00 & 63&7.54 & 24&7.28\\
  N II&3955.85 & --&  -- & --&  -- & --&  -- & --&  -- & --&  --\\
  N II&3995.00 &$<$13&$<$6.99 &$<$25&$<$6.88 & 64&7.58 &$<$21&$<$6.59 & $<$16&$<$6.88\\
  N II&4447.03 & --&  -- & --&  -- & 25&7.49 & --&  -- & --&  --\\
  N II&4601.48 & --&  -- & --&  -- & 26&7.67 & --&  -- & --&  --\\
\end{tabular}
\end{table*}

\setcounter{table}{0}
\clearpage
\begin{table*}
\caption[]{-continued.}
\begin{tabular}{lccccccccccc}
\hline \hline
Species&Wavelength&\multicolumn{2}{c}{NGC346-043}&\multicolumn{2}{c}{NGC346-044}
&\multicolumn{2}{c}{NGC346-047}&\multicolumn{2}{c}{NGC346-053}
&\multicolumn{2}{c}{NGC346-054}\\
       &          &EW&Abund&EW&Abund&EW&Abund&EW&Abund&EW&Abund\\
\hline \\
  C II&4267.00 & 15&7.37 & 84&7.50 & 92&7.18 & 58&7.77 & 32&7.29\\
  N II&3955.85 & --&  -- & --&  -- & --&  -- & --&  -- & --&  --\\
  N II&3995.00 & $<$7&$<$6.75 & $<$42&$<$6.83 & 55&7.29 & $<$45&$<$7.44 & $<$20&$<$6.77\\
  N II&4447.03 & --&  -- & $<$24&$<$7.05 & 26&7.36 & --&  -- & --&  --\\
  N II&4601.48 & --&  -- & --&  -- & 25&7.49 & --&  -- & --&  --\\
\end{tabular}
\end{table*}

\setcounter{table}{0}
\clearpage
\begin{table*}
\caption[]{-continued.}
\begin{tabular}{lccccccccccc}
\hline \hline
Species&Wavelength&\multicolumn{2}{c}{NGC346-055}&\multicolumn{2}{c}{NGC346-056}
&\multicolumn{2}{c}{NGC346-057}&\multicolumn{2}{c}{NGC346-058}
&\multicolumn{2}{c}{NGC346-062}\\
       &          &EW&Abund&EW&Abund&EW&Abund&EW&Abund&EW&Abund\\
\hline \\
  C II&4267.00 & 40&7.45 & 14&7.15 & 91&7.16 & --&  -- & 32&7.35\\
  N II&3955.85 & --&  -- & --&  -- & --&  -- & --&  -- & --&  --\\
  N II&3995.00 & $<$40&$<$7.25 & 33&7.47 & $<$25&$<$6.71 & $<$70&$<$7.56 & 36&7.20\\
  N II&4447.03 & --&  -- & 15&7.31 & --&  -- & --&  -- & --&  --\\
  N II&4601.48 & --&  -- & --&  -- & --&  -- & --&  -- & --&  --\\
\end{tabular}
\end{table*}

\setcounter{table}{0}
\clearpage
\begin{table*}
\caption[]{-continued.}
\begin{tabular}{lccccccccccc}
\hline \hline
Species&Wavelength&\multicolumn{2}{c}{NGC346-070}&\multicolumn{2}{c}{NGC346-074}
&\multicolumn{2}{c}{NGC346-075}&\multicolumn{2}{c}{NGC346-079}
&\multicolumn{2}{c}{NGC346-080}\\
       &          &EW&Abund&EW&Abund&EW&Abund&EW&Abund&EW&Abund\\
\hline \\
  C II&4267.00 & --&  -- & 59&6.92 & 59&7.56 & --&  -- & --&  --\\
  N II&3955.85 & --&  -- & --&  -- & --&  -- & --&  -- & --&  --\\
  N II&3995.00 & $<$50&$<$7.50 & 29&7.43 & $<$15&$<$6.48 & $<$90&$<$7.88 &$<$100&$<$7.85\\
  N II&4447.03 & --&  -- & --&  -- & --&  -- & --&  -- & --&  --\\
  N II&4601.48 & --&  -- & --&  -- & --&  -- & --&  -- & --&  --\\
\end{tabular}
\end{table*}

\setcounter{table}{0}
\clearpage
\begin{table*}
\caption[]{-continued.}
\begin{tabular}{lccccccccccc}
\hline \hline
Species&Wavelength&\multicolumn{2}{c}{NGC346-081}&\multicolumn{2}{c}{NGC346-082}
&\multicolumn{2}{c}{NGC346-083}&\multicolumn{2}{c}{NGC346-084}
&\multicolumn{2}{c}{NGC346-088}\\
       &          &EW&Abund&EW&Abund&EW&Abund&EW&Abund&EW&Abund\\
\hline \\
  C II&4267.00 & 55&7.02 & 87&7.30 & --&  -- & 54&7.48 & 80&7.78\\
  N II&3955.85 & --&  -- & --&  -- & --&  -- & --&  -- & --&  --\\
  N II&3995.00 & $<$80&$<$7.52 & $<$65&$<$7.55 & $<$90&$<$7.73 & $<$40&$<$7.06 & $<$35&$<$6.94\\
  N II&4447.03 & --&  -- & --&  -- & --&  -- & --&  -- & --&  --\\
  N II&4601.48 & --&  -- & --&  -- & --&  -- & --&  -- & --&  --\\
\end{tabular}
\end{table*}

\setcounter{table}{0}
\clearpage
\begin{table*}
\caption[]{-continued.}
\begin{tabular}{lccccccccccc}
\hline \hline
Species&Wavelength&\multicolumn{2}{c}{NGC346-092}&\multicolumn{2}{c}{NGC346-094}
&\multicolumn{2}{c}{NGC346-098}&\multicolumn{2}{c}{NGC346-099}
&\multicolumn{2}{c}{NGC346-100}\\
       &          &EW&Abund&EW&Abund&EW&Abund&EW&Abund&EW&Abund\\
\hline \\
  C II&4267.00 & 73&7.72 & 63&7.68 & 76&7.63 & 91&7.13 & 72&7.56\\
  N II&3955.85 & --&  -- & --&  -- & --&  -- & --&  -- & --&  --\\
  N II&3995.00 & $<$80&$<$7.52 & 57&7.34 & 38&6.94 & $<$30&$<$7.14 & $<$65&$<$7.41\\
  N II&4447.03 & --&  -- & --&  -- & --&  -- & --&  -- & --&  --\\
  N II&4601.48 & --&  -- & --&  -- & --&  -- & --&  -- & --&  --\\
\end{tabular}
\end{table*}

\setcounter{table}{0}
\clearpage
\begin{table*}
\caption[]{-continued.}
\begin{tabular}{lccccccccccc}
\hline \hline
Species&Wavelength&\multicolumn{2}{c}{NGC346-101}&\multicolumn{2}{c}{NGC346-102}
&\multicolumn{2}{c}{NGC346-103}&\multicolumn{2}{c}{NGC346-104}
&\multicolumn{2}{c}{NGC346-106}\\
       &          &EW&Abund&EW&Abund&EW&Abund&EW&Abund&EW&Abund\\
\hline \\
  C II&4267.00 & 42&7.33 & 50&6.88 & 26&7.23 & --&  -- & --&  --\\
  N II&3955.85 & --&  -- & --&  -- & --&  -- & --&  -- & --&  --\\
  N II&3995.00 & 20&6.63 & $<$15&$<$6.97 & 53&7.48 & $<$20&$<$7.20 & $<$70&$<$7.51\\
  N II&4447.03 & --&  -- & --&  -- & 43&7.75 & --&  -- & --&  --\\
  N II&4601.48 & --&  -- & --&  -- & 19&7.63 & --&  -- & --&  --\\
\end{tabular}
\end{table*}

\setcounter{table}{0}
\clearpage
\begin{table*}
\caption[]{-continued.}
\begin{tabular}{lccccccccccc}
\hline \hline
Species&Wavelength&\multicolumn{2}{c}{NGC346-108}&\multicolumn{2}{c}{NGC346-109}
&\multicolumn{2}{c}{NGC346-114}&\multicolumn{2}{c}{NGC346-116}
&\multicolumn{2}{c}{NGC330-002}\\
       &          &EW&Abund&EW&Abund&EW&Abund&EW&Abund&EW&Abund\\
\hline \\
  C II&4267.00 & 70&7.59 & 89&7.70 & --&  -- & 43&7.43 & 82&6.91\\
  N II&3955.85 & --&  -- & --&  -- & --&  -- & --&  -- & --&  --\\
  N II&3995.00 & $<$80&$<$7.42 & $<$50&$<$7.09 &$<$110&$<$7.80 & 28&6.93 & 56&7.49\\
  N II&4447.03 & --&  -- & --&  -- & --&  -- & --&  -- & 23&7.70\\
  N II&4601.48 & --&  -- & --&  -- & --&  -- & --&  -- & 18&7.62\\
\end{tabular}
\end{table*}

\setcounter{table}{0}
\clearpage
\begin{table*}
\caption[]{-continued.}
\begin{tabular}{lccccccccccc}
\hline \hline
Species&Wavelength&\multicolumn{2}{c}{NGC330-003}&\multicolumn{2}{c}{NGC330-004}
&\multicolumn{2}{c}{NGC330-005}&\multicolumn{2}{c}{NGC330-009}
&\multicolumn{2}{c}{NGC330-010}\\
       &          &EW&Abund&EW&Abund&EW&Abund&EW&Abund&EW&Abund\\
\hline \\
  C II&4267.00 &134&7.14 & 62&6.72 & 32&6.59 & 48&6.87 & 64&6.99\\
  N II&3955.85 & --&  -- & --&  -- & --&  -- & --&  -- & --&  --\\
  N II&3995.00 &150&7.62 &146&7.75 & 24&7.41 & 16&7.19 & 15&7.06\\
  N II&4447.03 & 67&7.74 & 64&7.80 & --&  -- & --&  -- & --&  --\\
  N II&4601.48 & 51&7.67 & 63&7.86 &  6&7.51 & --&  -- & --&  --\\
\end{tabular}
\end{table*}

\setcounter{table}{0}
\clearpage
\begin{table*}
\caption[]{-continued.}
\begin{tabular}{lccccccccccc}
\hline \hline
Species&Wavelength&\multicolumn{2}{c}{NGC330-014}&\multicolumn{2}{c}{NGC330-016}
&\multicolumn{2}{c}{NGC330-017}&\multicolumn{2}{c}{NGC330-018}
&\multicolumn{2}{c}{NGC330-020}\\
       &          &EW&Abund&EW&Abund&EW&Abund&EW&Abund&EW&Abund\\
\hline \\
  C II&4267.00 & 66&7.02 & 69&7.18 & 74&7.30 &102&7.10 & 76&6.92\\
  N II&3955.85 & --&  -- & --&  -- & --&  -- & --&  -- & --&  --\\
  N II&3995.00 &157&7.47 & 15&7.18 & 48&7.04 & 50&7.31 & 22&6.93\\
  N II&4447.03 & 69&7.54 & --&  -- & 24&7.12 & --&  -- & --&  --\\
  N II&4601.48 & 64&7.61 & --&  -- & 20&7.18 & --&  -- & --&  --\\
\end{tabular}
\end{table*}

\setcounter{table}{0}
\clearpage
\begin{table*}
\caption[]{-continued.}
\begin{tabular}{lccccccccccc}
\hline \hline
Species&Wavelength&\multicolumn{2}{c}{NGC330-021}&\multicolumn{2}{c}{NGC330-022}
&\multicolumn{2}{c}{NGC330-024}&\multicolumn{2}{c}{NGC330-026}
&\multicolumn{2}{c}{NGC330-027}\\
       &          &EW&Abund&EW&Abund&EW&Abund&EW&Abund&EW&Abund\\
\hline \\
  C II&4267.00 & --&  -- & 94&7.11 & 53&7.09 & 97&7.57 & 81&7.36\\
  N II&3955.85 & --&  -- & --&  -- & --&  -- & --&  -- & --&  --\\
  N II&3995.00 & --&  -- & 50&7.30 & $<$20&$<$7.53 & 77&7.45 &100&7.38\\
  N II&4447.03 & --&  -- & 16&7.12 & --&  -- & --&  -- & --&  --\\
  N II&4601.48 & --&  -- & 17&7.30 & --&  -- & 22&7.22 & --&  --\\
\end{tabular}
\end{table*}

\setcounter{table}{0}
\clearpage
\begin{table*}
\caption[]{-continued.}
\begin{tabular}{lccccccccccc}
\hline \hline
Species&Wavelength&\multicolumn{2}{c}{NGC330-028}&\multicolumn{2}{c}{NGC330-032}
&\multicolumn{2}{c}{NGC330-033}&\multicolumn{2}{c}{NGC330-035}
&\multicolumn{2}{c}{NGC330-036}\\
       &          &EW&Abund&EW&Abund&EW&Abund&EW&Abund&EW&Abund\\
\hline \\
  C II&4267.00 & 88&7.84 & 33&7.37 & 86&7.74 & 70&6.94 & 75&7.18\\
  N II&3955.85 & --&  -- & --&  -- & --&  -- & --&  -- & --&  --\\
  N II&3995.00 & 76&7.44 & 43&7.33 & 44&7.07 & 39&7.19 & 65&7.12\\
  N II&4447.03 & --&  -- & 24&7.38 & --&  -- & --&  -- & --&  --\\
  N II&4601.48 & --&  -- & --&  -- & --&  -- & --&  -- & --&  --\\
\end{tabular}
\end{table*}

\setcounter{table}{0}
\clearpage
\begin{table*}
\caption[]{-continued.}
\begin{tabular}{lccccccccccc}
\hline \hline
Species&Wavelength&\multicolumn{2}{c}{NGC330-038}&\multicolumn{2}{c}{NGC330-039}
&\multicolumn{2}{c}{NGC330-040}&\multicolumn{2}{c}{NGC330-041}
&\multicolumn{2}{c}{NGC330-042}\\
       &          &EW&Abund&EW&Abund&EW&Abund&EW&Abund&EW&Abund\\
\hline \\
  C II&4267.00 & 66&7.67 & --&  -- &105&7.39 & 53&7.86 & 48&7.33\\
  N II&3955.85 & --&  -- & --&  -- & --&  -- & --&  -- & --&  --\\
  N II&3995.00 & $<$45&$<$7.16 & $<$35&$<$7.61 & --&  -- & --&  -- & 59&7.22\\
  N II&4447.03 & --&  -- & --&  -- & --&  -- & --&  -- & 28&7.21\\
  N II&4601.48 & --&  -- & --&  -- & --&  -- & --&  -- & 16&7.14\\
\end{tabular}
\end{table*}

\setcounter{table}{0}
\clearpage
\begin{table*}
\caption[]{-continued.}
\begin{tabular}{lccccccccccc}
\hline \hline
Species&Wavelength&\multicolumn{2}{c}{NGC330-045}&\multicolumn{2}{c}{NGC330-047}
&\multicolumn{2}{c}{NGC330-048}&\multicolumn{2}{c}{NGC330-049}
&\multicolumn{2}{c}{NGC330-051}\\
       &          &EW&Abund&EW&Abund&EW&Abund&EW&Abund&EW&Abund\\
\hline \\
  C II&4267.00 & 77&6.96 & 54&7.47 & 39&7.42 & --&  -- & --&  --\\
  N II&3955.85 & --&  -- & --&  -- & --&  -- & --&  -- & --&  --\\
  N II&3995.00 & $<$50&$<$7.28 & 20&6.61 & $<$35&$<$7.10 & 25&6.69 & $<$60&$<$7.20\\
  N II&4447.03 & --&  -- & --&  -- & --&  -- & --&  -- & --&  --\\
  N II&4601.48 & --&  -- & --&  -- & --&  -- & --&  -- & --&  --\\
\end{tabular}
\end{table*}

\setcounter{table}{0}
\clearpage
\begin{table*}
\caption[]{-continued.}
\begin{tabular}{lccccccccccc}
\hline \hline
Species&Wavelength&\multicolumn{2}{c}{NGC330-053}&\multicolumn{2}{c}{NGC330-055}
&\multicolumn{2}{c}{NGC330-056}&\multicolumn{2}{c}{NGC330-057}
&\multicolumn{2}{c}{NGC330-059}\\
       &          &EW&Abund&EW&Abund&EW&Abund&EW&Abund&EW&Abund\\
\hline \\
  C II&4267.00 & 48&7.54 & --&  -- & 84&7.25 & --&  -- &103&7.17\\
  N II&3955.85 & --&  -- & --&  -- & --&  -- & --&  -- & --&  --\\
  N II&3995.00 & $<$30&$<$7.04 & 90&8.06 & $<$45&$<$6.93 & $<$65&$<$7.48 & 55&7.27\\
  N II&4447.03 & --&  -- & --&  -- & --&  -- & --&  -- & --&  --\\
  N II&4601.48 & --&  -- & --&  -- & --&  -- & --&  -- & --&  --\\
\end{tabular}
\end{table*}

\setcounter{table}{0}
\clearpage
\begin{table*}
\caption[]{-continued.}
\begin{tabular}{lccccccccccc}
\hline \hline
Species&Wavelength&\multicolumn{2}{c}{NGC330-066}&\multicolumn{2}{c}{NGC330-067}
&\multicolumn{2}{c}{NGC330-069}&\multicolumn{2}{c}{NGC330-072}
&\multicolumn{2}{c}{NGC330-074}\\
       &          &EW&Abund&EW&Abund&EW&Abund&EW&Abund&EW&Abund\\
\hline \\
  C II&4267.00 & 84&7.08 & 92&7.15 &114&7.30 & 58&7.67 & 23&7.42\\
  N II&3955.85 & --&  -- & --&  -- & --&  -- & --&  -- & --&  --\\
  N II&3995.00 & $<$70&$<$7.77 & 55&7.17 & $<$90&$<$7.91 & $<$45&$<$7.30 & 25&7.26\\
  N II&4447.03 & --&  -- & --&  -- & --&  -- & --&  -- & --&  --\\
  N II&4601.48 & --&  -- & --&  -- & --&  -- & --&  -- & --&  --\\
\end{tabular}
\end{table*}

\setcounter{table}{0}
\clearpage
\begin{table*}
\caption[]{-continued.}
\begin{tabular}{lccccccccccc}
\hline \hline
Species&Wavelength&\multicolumn{2}{c}{NGC330-079}&\multicolumn{2}{c}{NGC330-083}
&\multicolumn{2}{c}{NGC330-086}&\multicolumn{2}{c}{NGC330-095}
&\multicolumn{2}{c}{NGC330-097}\\
       &          &EW&Abund&EW&Abund&EW&Abund&EW&Abund&EW&Abund\\
\hline \\
  C II&4267.00 &111&7.29 & 91&7.09 & 76&7.06 &107&7.25 & 80&7.76\\
  N II&3955.85 & --&  -- & --&  -- & --&  -- & --&  -- & --&  --\\
  N II&3995.00 & $<$80&$<$7.69 & $<$60&$<$7.66 & $<$60&$<$7.33 & $<$25&$<$6.95 & 88&7.53\\
  N II&4447.03 & --&  -- & --&  -- & --&  -- & --&  -- & --&  --\\
  N II&4601.48 & --&  -- & --&  -- & --&  -- & --&  -- & --&  --\\
\end{tabular}
\end{table*}

\setcounter{table}{0}
\clearpage
\begin{table*}
\caption[]{-continued.}
\begin{tabular}{lccccccccccc}
\hline \hline
Species&Wavelength&\multicolumn{2}{c}{NGC330-101}&\multicolumn{2}{c}{NGC330-107}
&\multicolumn{2}{c}{NGC330-108}&\multicolumn{2}{c}{NGC330-109}
&\multicolumn{2}{c}{NGC330-110}\\
       &          &EW&Abund&EW&Abund&EW&Abund&EW&Abund&EW&Abund\\
\hline \\
  C II&4267.00 &102&7.27 & 82&7.16 & --&  -- & --&  -- & 62&7.10\\
  N II&3955.85 & --&  -- & --&  -- & --&  -- & --&  -- & --&  --\\
  N II&3995.00 & 37&7.07 & $<$30&$<$7.42 & --&  -- & --&  -- & 60&7.03\\
  N II&4447.03 & --&  -- & --&  -- & --&  -- & --&  -- & --&  --\\
  N II&4601.48 & --&  -- & --&  -- & --&  -- & --&  -- & --&  --\\
\end{tabular}
\end{table*}

\setcounter{table}{0}
\clearpage
\begin{table*}
\caption[]{-continued.}
\begin{tabular}{lccccccccccc}
\hline \hline
Species&Wavelength&\multicolumn{2}{c}{NGC330-114}&\multicolumn{2}{c}{NGC330-116}
&\multicolumn{2}{c}{NGC330-120}&\multicolumn{2}{c}{NGC330-123}
&\multicolumn{2}{c}{NGC330-124}\\
       &          &EW&Abund&EW&Abund&EW&Abund&EW&Abund&EW&Abund\\
\hline \\
  C II&4267.00 & 83&7.53 &118&7.37 & 64&6.90 & --&  -- & 54&7.71\\
  N II&3955.85 & --&  -- & --&  -- & --&  -- & --&  -- & --&  --\\
  N II&3995.00 & 53&7.08 & 34&7.05 & $<$76&$<$7.85 & --&  -- & 38&7.36\\
  N II&4447.03 & --&  -- & --&  -- & --&  -- & --&  -- & --&  --\\
  N II&4601.48 & --&  -- & --&  -- & --&  -- & --&  -- & --&  --\\
\end{tabular}
\end{table*}

\setcounter{table}{0}
\clearpage
\begin{table*}
\caption[]{-continued.}
\begin{tabular}{lccccccccccc}
\hline \hline
Species&Wavelength&\multicolumn{2}{c}{NGC330-125}&\multicolumn{2}{c}{N11-001}
&\multicolumn{2}{c}{N11-002}&\multicolumn{2}{c}{N11-003}
&\multicolumn{2}{c}{N11-008}\\
       &          &EW&Abund&EW&Abund&EW&Abund&EW&Abund&EW&Abund\\
\hline \\
  C II&4267.00 & 68&7.14 &146&7.32 &190&7.38 & 79&7.64 & 84&7.76\\
  N II&3955.85 & --&  -- & --&  -- & 42&8.01 & --&  -- & --&  --\\
  N II&3995.00 & 70&7.28 &269&8.20 &164&7.95 & 48&7.09 &122&7.79\\
  N II&4447.03 & --&  -- &111&7.96 & 66&7.91 & --&  -- & --&  --\\
  N II&4601.48 & --&  -- &137&8.24 & 67&7.99 & --&  -- & --&  --\\
\end{tabular}
\end{table*}

\setcounter{table}{0}
\clearpage
\begin{table*}
\caption[]{-continued.}
\begin{tabular}{lccccccccccc}
\hline \hline
Species&Wavelength&\multicolumn{2}{c}{N11-009}&\multicolumn{2}{c}{N11-012}
&\multicolumn{2}{c}{N11-014}&\multicolumn{2}{c}{N11-015}
&\multicolumn{2}{c}{N11-016}\\
       &          &EW&Abund&EW&Abund&EW&Abund&EW&Abund&EW&Abund\\
\hline \\
  C II&4267.00 &187&7.41 & 94&7.35 &188&7.59 & 97&7.75 &135&7.76\\
  N II&3955.85 & --&  -- & --&  -- & --&  -- & --&  -- & --&  --\\
  N II&3995.00 & 89&7.68 &173&7.72 &202&7.93 & 60&7.14 &217&7.93\\
  N II&4447.03 & 35&7.76 & --&  -- & 74&7.69 & --&  -- & --&  --\\
  N II&4601.48 & 28&7.70 & 68&7.70 & 92&7.94 & --&  -- &107&7.99\\
\end{tabular}
\end{table*}

\setcounter{table}{0}
\clearpage
\begin{table*}
\caption[]{-continued.}
\begin{tabular}{lccccccccccc}
\hline \hline
Species&Wavelength&\multicolumn{2}{c}{N11-017}&\multicolumn{2}{c}{N11-023}
&\multicolumn{2}{c}{N11-024}&\multicolumn{2}{c}{N11-029}
&\multicolumn{2}{c}{N11-034}\\
       &          &EW&Abund&EW&Abund&EW&Abund&EW&Abund&EW&Abund\\
\hline \\
  C II&4267.00 &197&7.38 & 92&7.76 &123&7.65 & 55&7.81 &142&8.10\\
  N II&3955.85 & --&  -- & --&  -- & --&  -- & --&  -- & --&  --\\
  N II&3995.00 &143&7.79 & 54&7.16 &196&7.80 & 19&7.10 & 61&7.22\\
  N II&4447.03 & 59&7.79 & --&  -- & --&  -- & --&  -- & --&  --\\
  N II&4601.48 & 60&7.87 & --&  -- & 98&7.90 & --&  -- & --&  --\\
\end{tabular}
\end{table*}

\setcounter{table}{0}
\clearpage
\begin{table*}
\caption[]{-continued.}
\begin{tabular}{lccccccccccc}
\hline \hline
Species&Wavelength&\multicolumn{2}{c}{N11-036}&\multicolumn{2}{c}{N11-037}
&\multicolumn{2}{c}{N11-039}&\multicolumn{2}{c}{N11-042}
&\multicolumn{2}{c}{N11-046}\\
       &          &EW&Abund&EW&Abund&EW&Abund&EW&Abund&EW&Abund\\
\hline \\
  C II&4267.00 & 91&7.66 & 65&7.81 &202&8.02 & 61&7.75 & 37&7.90\\
  N II&3955.85 & --&  -- & --&  -- & --&  -- & --&  -- & --&  --\\
  N II&3995.00 &154&7.74 & $<$27&$<$7.17 & 53&6.89 & 18&6.92 & $<$31&$<$7.71\\
  N II&4447.03 & 80&7.74 & --&  -- & --&  -- & --&  -- & --&  --\\
  N II&4601.48 & --&  -- & --&  -- & --&  -- & --&  -- & --&  --\\
\end{tabular}
\end{table*}

\setcounter{table}{0}
\clearpage
\begin{table*}
\caption[]{-continued.}
\begin{tabular}{lccccccccccc}
\hline \hline
Species&Wavelength&\multicolumn{2}{c}{N11-047}&\multicolumn{2}{c}{N11-054}
&\multicolumn{2}{c}{N11-059}&\multicolumn{2}{c}{N11-062}
&\multicolumn{2}{c}{N11-069}\\
       &          &EW&Abund&EW&Abund&EW&Abund&EW&Abund&EW&Abund\\
\hline \\
  C II&4267.00 & 77&7.89 &115&7.83 & --&  -- & 48&7.61 &131&7.92\\
  N II&3955.85 & --&  -- & --&  -- & --&  -- & --&  -- & --&  --\\
  N II&3995.00 & $<$16&$<$6.88 & 33&6.79 & --&  -- & 30&7.20 & 46&6.94\\
  N II&4447.03 & --&  -- & --&  -- & --&  -- & 14&7.12 & --&  --\\
  N II&4601.48 & --&  -- & --&  -- & --&  -- & --&  -- & --&  --\\
\end{tabular}
\end{table*}

\setcounter{table}{0}
\clearpage
\begin{table*}
\caption[]{-continued.}
\begin{tabular}{lccccccccccc}
\hline \hline
Species&Wavelength&\multicolumn{2}{c}{N11-070}&\multicolumn{2}{c}{N11-072}
&\multicolumn{2}{c}{N11-075}&\multicolumn{2}{c}{N11-077}
&\multicolumn{2}{c}{N11-079}\\
       &          &EW&Abund&EW&Abund&EW&Abund&EW&Abund&EW&Abund\\
\hline \\
  C II&4267.00 &132&7.48 & 74&7.82 &147&7.74 &139&7.66 & 43&7.74\\
  N II&3955.85 & --&  -- & --&  -- & --&  -- & --&  -- & --&  --\\
  N II&3995.00 & 24&6.75 & 52&7.41 &133&7.84 & 46&6.92 & 13&7.01\\
  N II&4447.03 & --&  -- & 25&7.35 & 76&7.82 & --&  -- & --&  --\\
  N II&4601.48 & --&  -- & --&  -- & 85&8.16 & --&  -- & --&  --\\
\end{tabular}
\end{table*}

\setcounter{table}{0}
\clearpage
\begin{table*}
\caption[]{-continued.}
\begin{tabular}{lccccccccccc}
\hline \hline
Species&Wavelength&\multicolumn{2}{c}{N11-083}&\multicolumn{2}{c}{N11-084}
&\multicolumn{2}{c}{N11-086}&\multicolumn{2}{c}{N11-088}
&\multicolumn{2}{c}{N11-089}\\
       &          &EW&Abund&EW&Abund&EW&Abund&EW&Abund&EW&Abund\\
\hline \\
  C II&4267.00 & 76&7.85 & 69&7.76 & 48&7.36 & 84&7.59 &168&7.88\\
  N II&3955.85 & --&  -- & --&  -- & --&  -- & --&  -- & --&  --\\
  N II&3995.00 & 19&6.86 & $<$21&$<$6.89 & 28&6.84 & $<$49&$<$6.96 & 47&6.88\\
  N II&4447.03 & --&  -- & --&  -- & --&  -- & --&  -- & --&  --\\
  N II&4601.48 & --&  -- & --&  -- & --&  -- & --&  -- & --&  --\\
\end{tabular}
\end{table*}

\setcounter{table}{0}
\clearpage
\begin{table*}
\caption[]{-continued.}
\begin{tabular}{lccccccccccc}
\hline \hline
Species&Wavelength&\multicolumn{2}{c}{N11-093}&\multicolumn{2}{c}{N11-095}
&\multicolumn{2}{c}{N11-097}&\multicolumn{2}{c}{N11-098}
&\multicolumn{2}{c}{N11-100}\\
       &          &EW&Abund&EW&Abund&EW&Abund&EW&Abund&EW&Abund\\
\hline \\
  C II&4267.00 &150&7.60 & 69&7.63 &116&7.31 &140&7.71 & 56&7.65\\
  N II&3955.85 & --&  -- & --&  -- & --&  -- & --&  -- & --&  --\\
  N II&3995.00 & 20&6.65 & 70&7.46 & 48&7.53 & 48&7.00 & 80&7.77\\
  N II&4447.03 & --&  -- & --&  -- & 29&7.77 & --&  -- & 29&7.47\\
  N II&4601.48 & --&  -- & --&  -- & --&  -- & --&  -- & --&  --\\
\end{tabular}
\end{table*}

\setcounter{table}{0}
\clearpage
\begin{table*}
\caption[]{-continued.}
\begin{tabular}{lccccccccccc}
\hline \hline
Species&Wavelength&\multicolumn{2}{c}{N11-101}&\multicolumn{2}{c}{N11-102}
&\multicolumn{2}{c}{N11-104}&\multicolumn{2}{c}{N11-106}
&\multicolumn{2}{c}{N11-108}\\
       &          &EW&Abund&EW&Abund&EW&Abund&EW&Abund&EW&Abund\\
\hline \\
  C II&4267.00 & 88&7.94 & 60&7.78 &114&7.96 & 44&7.66 & 48&7.83\\
  N II&3955.85 & --&  -- & --&  -- & --&  -- & --&  -- & --&  --\\
  N II&3995.00 & $<$28&$<$7.09 & $<$48&$<$7.58 & --&  -- & 20&7.13 & 18&7.21\\
  N II&4447.03 & --&  -- & --&  -- & --&  -- & --&  -- & --&  --\\
  N II&4601.48 & --&  -- & --&  -- & --&  -- & --&  -- & --&  --\\
\end{tabular}
\end{table*}

\setcounter{table}{0}
\clearpage
\begin{table*}
\caption[]{-continued.}
\begin{tabular}{lccccccccccc}
\hline \hline
Species&Wavelength&\multicolumn{2}{c}{N11-109}&\multicolumn{2}{c}{N11-110}
&\multicolumn{2}{c}{N11-111}&\multicolumn{2}{c}{N11-114}
&\multicolumn{2}{c}{N11-115}\\
       &          &EW&Abund&EW&Abund&EW&Abund&EW&Abund&EW&Abund\\
\hline \\
  C II&4267.00 & 82&7.75 &121&7.79 &108&7.67 & --&  -- & 89&7.63\\
  N II&3955.85 & --&  -- & --&  -- & --&  -- & --&  -- & --&  --\\
  N II&3995.00 & 55&7.24 &108&7.45 & 43&6.87 & $<$54&$<$7.92 & 34&6.82\\
  N II&4447.03 & --&  -- & 43&7.33 & --&  -- & --&  -- & --&  --\\
  N II&4601.48 & --&  -- & --&  -- & --&  -- & --&  -- & --&  --\\
\end{tabular}
\end{table*}

\setcounter{table}{0}
\clearpage
\begin{table*}
\caption[]{-continued.}
\begin{tabular}{lccccccccccc}
\hline \hline
Species&Wavelength&\multicolumn{2}{c}{N11-116}&\multicolumn{2}{c}{N11-117}
&\multicolumn{2}{c}{N11-118}&\multicolumn{2}{c}{N11-120}
&\multicolumn{2}{c}{N11-121}\\
       &          &EW&Abund&EW&Abund&EW&Abund&EW&Abund&EW&Abund\\
\hline \\
  C II&4267.00 &145&7.74 & 56&7.46 &113&7.93 & 62&7.90 & 57&7.49\\
  N II&3955.85 & --&  -- & --&  -- & --&  -- & --&  -- & --&  --\\
  N II&3995.00 & 42&6.86 & $<$40&$<$7.06 & 62&7.28 & $<$41&$<$7.56 & $<$67&$<$7.48\\
  N II&4447.03 & --&  -- & --&  -- & 38&7.41 & --&  -- & --&  --\\
  N II&4601.48 & --&  -- & --&  -- & --&  -- & --&  -- & --&  --\\
\end{tabular}
\end{table*}

\setcounter{table}{0}
\clearpage
\begin{table*}
\caption[]{-continued.}
\begin{tabular}{lccccccccccc}
\hline \hline
Species&Wavelength&\multicolumn{2}{c}{N11-122}&\multicolumn{2}{c}{N11-124}
&\multicolumn{2}{c}{NGC2004-003}&\multicolumn{2}{c}{NGC2004-010}
&\multicolumn{2}{c}{NGC2004-011}\\
       &          &EW&Abund&EW&Abund&EW&Abund&EW&Abund&EW&Abund\\
\hline \\
  C II&4267.00 & --&  -- & 78&7.80 &173&7.48 &194&7.42 &130&7.63\\
  N II&3955.85 & --&  -- & --&  -- & 21&7.89 & 64&8.19 & 52&7.85\\
  N II&3995.00 & $<$34&$<$7.72 & 43&7.25 & 83&7.79 &202&8.11 &196&7.76\\
  N II&4447.03 & --&  -- & --&  -- & 31&7.81 & 93&8.04 & 76&7.60\\
  N II&4601.48 & --&  -- & --&  -- & 31&7.86 & 97&8.17 & 95&7.85\\
\end{tabular}
\end{table*}

\setcounter{table}{0}
\clearpage
\begin{table*}
\caption[]{-continued.}
\begin{tabular}{lccccccccccc}
\hline \hline
Species&Wavelength&\multicolumn{2}{c}{NGC2004-012}
&\multicolumn{2}{c}{NGC2004-013}&\multicolumn{2}{c}{NGC2004-014}
&\multicolumn{2}{c}{NGC2004-015}&\multicolumn{2}{c}{NGC2004-020}\\
       &          &EW&Abund&EW&Abund&EW&Abund&EW&Abund&EW&Abund\\
\hline \\
  C II&4267.00 &134&7.62 &146&7.77 &191&7.50 &176&8.05 &187&8.14\\
  N II&3955.85 & 48&7.80 & --&  -- & 19&7.56 & --&  -- & --&  --\\
  N II&3995.00 &177&7.69 & 42&6.84 & 82&7.47 & 58&6.97 &105&7.37\\
  N II&4447.03 & 69&7.54 & --&  -- & 33&7.49 & --&  -- & 72&7.60\\
  N II&4601.48 &105&7.94 & --&  -- & 30&7.55 & --&  -- & --&  --\\
\end{tabular}
\end{table*}

\setcounter{table}{0}
\clearpage
\begin{table*}
\caption[]{-continued.}
\begin{tabular}{lccccccccccc}
\hline \hline
Species&Wavelength&\multicolumn{2}{c}{NGC2004-021}
&\multicolumn{2}{c}{NGC2004-022}&\multicolumn{2}{c}{NGC2004-024}
&\multicolumn{2}{c}{NGC2004-026}&\multicolumn{2}{c}{NGC2004-029}\\
       &          &EW&Abund&EW&Abund&EW&Abund&EW&Abund&EW&Abund\\
\hline \\
  C II&4267.00 &167&7.74 &119&7.53 & 43&7.18 &131&7.82 &112&7.73\\
  N II&3955.85 & 18&7.26 & 38&7.70 & --&  -- & --&  -- & --&  --\\
  N II&3995.00 & 82&7.07 &149&7.58 &282&8.13 & --&  -- & 41&6.87\\
  N II&4447.03 & 22&6.93 & 68&7.54 &124&7.91 & --&  -- & 12&6.75\\
  N II&4601.48 & 42&7.37 & 88&7.86 & --&  -- & 21&7.20 & --&  --\\
\end{tabular}
\end{table*}

\setcounter{table}{0}
\clearpage
\begin{table*}
\caption[]{-continued.}
\begin{tabular}{lccccccccccc}
\hline \hline
Species&Wavelength&\multicolumn{2}{c}{NGC2004-030}
&\multicolumn{2}{c}{NGC2004-031}&\multicolumn{2}{c}{NGC2004-032}
&\multicolumn{2}{c}{NGC2004-036}&\multicolumn{2}{c}{NGC2004-041}\\
       &          &EW&Abund&EW&Abund&EW&Abund&EW&Abund&EW&Abund\\
\hline \\
  C II&4267.00 & 87&7.92 &159&7.81 &149&7.73 &131&7.80 &156&7.67\\
  N II&3955.85 & --&  -- & --&  -- & --&  -- & 22&7.46 & --&  --\\
  N II&3995.00 & --&  -- & 41&6.84 & 41&6.84 & 88&7.28 & 32&6.80\\
  N II&4447.03 & 30&7.43 & --&  -- & --&  -- & 41&7.30 & --&  --\\
  N II&4601.48 & --&  -- & --&  -- & --&  -- & --&  -- & --&  --\\
\end{tabular}
\end{table*}

\setcounter{table}{0}
\clearpage
\begin{table*}
\caption[]{-continued.}
\begin{tabular}{lccccccccccc}
\hline \hline
Species&Wavelength&\multicolumn{2}{c}{NGC2004-042}
&\multicolumn{2}{c}{NGC2004-043}&\multicolumn{2}{c}{NGC2004-045}
&\multicolumn{2}{c}{NGC2004-046}&\multicolumn{2}{c}{NGC2004-047}\\
       &          &EW&Abund&EW&Abund&EW&Abund&EW&Abund&EW&Abund\\
\hline \\
  C II&4267.00 &165&7.80 &131&7.78 &119&7.58 & 95&7.80 &120&7.56\\
  N II&3955.85 & --&  -- & 16&7.31 & --&  -- & 13&7.44 & --&  --\\
  N II&3995.00 & 37&6.88 & 73&7.14 & 38&6.89 & 81&7.54 & 46&6.91\\
  N II&4447.03 & --&  -- & 31&7.14 & --&  -- & 42&7.49 & --&  --\\
  N II&4601.48 & --&  -- & --&  -- & --&  -- & --&  -- & --&  --\\
\end{tabular}
\end{table*}

\setcounter{table}{0}
\clearpage
\begin{table*}
\caption[]{-continued.}
\begin{tabular}{lccccccccccc}
\hline \hline
Species&Wavelength&\multicolumn{2}{c}{NGC2004-050}
&\multicolumn{2}{c}{NGC2004-051}&\multicolumn{2}{c}{NGC2004-052}
&\multicolumn{2}{c}{NGC2004-053}&\multicolumn{2}{c}{NGC2004-054}\\
       &          &EW&Abund&EW&Abund&EW&Abund&EW&Abund&EW&Abund\\
\hline \\
  C II&4267.00 &141&7.57 &150&7.73 &134&7.59 & 76&7.99 &147&7.78\\
  N II&3955.85 & --&  -- & --&  -- & --&  -- & --&  -- & --&  --\\
  N II&3995.00 & 50&7.11 & 42&6.85 & 42&6.92 & 50&7.63 & 46&6.98\\
  N II&4447.03 & --&  -- & --&  -- & --&  -- & --&  -- & --&  --\\
  N II&4601.48 & --&  -- & --&  -- & --&  -- & --&  -- & --&  --\\
\end{tabular}
\end{table*}

\setcounter{table}{0}
\clearpage
\begin{table*}
\caption[]{-continued.}
\begin{tabular}{lccccccccccc}
\hline \hline
Species&Wavelength&\multicolumn{2}{c}{NGC2004-059}
&\multicolumn{2}{c}{NGC2004-060}&\multicolumn{2}{c}{NGC2004-061}
&\multicolumn{2}{c}{NGC2004-062}&\multicolumn{2}{c}{NGC2004-063}\\
       &          &EW&Abund&EW&Abund&EW&Abund&EW&Abund&EW&Abund\\
\hline \\
  C II&4267.00 &154&7.82 &136&7.62 &175&7.91 & 66&7.89 &138&7.72\\
  N II&3955.85 & --&  -- & --&  -- & --&  -- & --&  -- & --&  --\\
  N II&3995.00 & 46&6.99 & 32&6.75 & 41&6.94 & $<$30&$<$7.33 & 43&6.98\\
  N II&4447.03 & --&  -- & --&  -- & 18&6.99 & --&  -- & --&  --\\
  N II&4601.48 & --&  -- & --&  -- & --&  -- & --&  -- & --&  --\\
\end{tabular}
\end{table*}

\setcounter{table}{0}
\clearpage
\begin{table*}
\caption[]{-continued.}
\begin{tabular}{lccccccccccc}
\hline \hline
Species&Wavelength&\multicolumn{2}{c}{NGC2004-064}
&\multicolumn{2}{c}{NGC2004-065}&\multicolumn{2}{c}{NGC2004-066}
&\multicolumn{2}{c}{NGC2004-068}&\multicolumn{2}{c}{NGC2004-069}\\
       &          &EW&Abund&EW&Abund&EW&Abund&EW&Abund&EW&Abund\\
\hline \\
  C II&4267.00 & 94&7.78 &155&7.70 & 92&7.78 &149&7.69 & 93&7.90\\
  N II&3955.85 & 17&7.55 & --&  -- & --&  -- & --&  -- & --&  --\\
  N II&3995.00 & 96&7.56 & $<$50&$<$7.19 & 92&7.64 & 35&7.01 & 89&7.68\\
  N II&4447.03 & 46&7.48 & --&  -- & --&  -- & --&  -- & --&  --\\
  N II&4601.48 & 31&7.50 & --&  -- & --&  -- & --&  -- & --&  --\\
\end{tabular}
\end{table*}

\setcounter{table}{0}
\clearpage
\begin{table*}
\caption[]{-continued.}
\begin{tabular}{lccccccccccc}
\hline \hline
Species&Wavelength&\multicolumn{2}{c}{NGC2004-070}
&\multicolumn{2}{c}{NGC2004-071}&\multicolumn{2}{c}{NGC2004-073}
&\multicolumn{2}{c}{NGC2004-074}&\multicolumn{2}{c}{NGC2004-075}\\
       &          &EW&Abund&EW&Abund&EW&Abund&EW&Abund&EW&Abund\\
\hline \\
  C II&4267.00 & 88&7.84 & 95&7.58 &146&7.93 & 87&7.78 &173&7.93\\
  N II&3955.85 & --&  -- & --&  -- & --&  -- & --&  -- & --&  --\\
  N II&3995.00 & 82&7.59 & 79&7.36 & 46&6.96 & --&  -- & 56&7.16\\
  N II&4447.03 & 34&7.42 & --&  -- & 20&6.98 & --&  -- & --&  --\\
  N II&4601.48 & --&  -- & --&  -- & --&  -- & --&  -- & --&  --\\
\end{tabular}
\end{table*}

\setcounter{table}{0}
\clearpage
\begin{table*}
\caption[]{-continued.}
\begin{tabular}{lccccccccccc}
\hline \hline
Species&Wavelength&\multicolumn{2}{c}{NGC2004-076}
&\multicolumn{2}{c}{NGC2004-077}&\multicolumn{2}{c}{NGC2004-078}
&\multicolumn{2}{c}{NGC2004-079}&\multicolumn{2}{c}{NGC2004-080}\\
       &          &EW&Abund&EW&Abund&EW&Abund&EW&Abund&EW&Abund\\
\hline \\
  C II&4267.00 &150&7.70 & 92&7.98 &125&7.61 &133&7.65 &144&7.64\\
  N II&3955.85 & --&  -- & --&  -- & --&  -- & --&  -- & --&  --\\
  N II&3995.00 & --&  -- & --&  -- & 30&6.74 &100&7.77 & 44&7.11\\
  N II&4447.03 & --&  -- & --&  -- & --&  -- & --&  -- & --&  --\\
  N II&4601.48 & --&  -- & --&  -- & --&  -- & --&  -- & --&  --\\
\end{tabular}
\end{table*}

\setcounter{table}{0}
\clearpage
\begin{table*}
\caption[]{-continued.}
\begin{tabular}{lccccccccccc}
\hline \hline
Species&Wavelength&\multicolumn{2}{c}{NGC2004-081}
&\multicolumn{2}{c}{NGC2004-082}&\multicolumn{2}{c}{NGC2004-084}
&\multicolumn{2}{c}{NGC2004-085}&\multicolumn{2}{c}{NGC2004-086}\\
       &          &EW&Abund&EW&Abund&EW&Abund&EW&Abund&EW&Abund\\
\hline \\
  C II&4267.00 &100&7.88 &132&8.03 & 96&7.90 &155&7.63 &140&7.65\\
  N II&3955.85 & --&  -- & --&  -- & --&  -- & --&  -- & --&  --\\
  N II&3995.00 & 42&7.07 & --&  -- & 35&6.99 & 55&7.18 & 51&7.11\\
  N II&4447.03 & --&  -- & --&  -- & --&  -- & --&  -- & 26&7.27\\
  N II&4601.48 & --&  -- & --&  -- & --&  -- & --&  -- & --&  --\\
\end{tabular}
\end{table*}

\setcounter{table}{0}
\clearpage
\begin{table*}
\caption[]{-continued.}
\begin{tabular}{lccccccccccc}
\hline \hline
Species&Wavelength&\multicolumn{2}{c}{NGC2004-087}
&\multicolumn{2}{c}{NGC2004-088}&\multicolumn{2}{c}{NGC2004-090}
&\multicolumn{2}{c}{NGC2004-091}&\multicolumn{2}{c}{NGC2004-094}\\
       &          &EW&Abund&EW&Abund&EW&Abund&EW&Abund&EW&Abund\\
\hline \\
  C II&4267.00 & 88&7.65 &166&7.65 & 41&7.79 &125&8.08 & 68&7.04\\
  N II&3955.85 & --&  -- & --&  -- & --&  -- & --&  -- & --&  --\\
  N II&3995.00 & 32&6.88 & $<$54&$<$7.18 & 40&7.72 & 50&7.21 & --&  --\\
  N II&4447.03 & --&  -- & --&  -- & 20&7.51 & 33&7.39 & --&  --\\
  N II&4601.48 & --&  -- & --&  -- & --&  -- & --&  -- & --&  --\\
\end{tabular}
\end{table*}

\setcounter{table}{0}
\clearpage
\begin{table*}
\caption[]{-continued.}
\begin{tabular}{lccccccccccc}
\hline \hline
Species&Wavelength&\multicolumn{2}{c}{NGC2004-095}
&\multicolumn{2}{c}{NGC2004-097}&\multicolumn{2}{c}{NGC2004-098}
&\multicolumn{2}{c}{NGC2004-099}&\multicolumn{2}{c}{NGC2004-100}\\
       &          &EW&Abund&EW&Abund&EW&Abund&EW&Abund&EW&Abund\\
\hline \\
  C II&4267.00 & 90&7.71 &141&7.71 &102&7.44 &125&7.61 & 95&7.85\\
  N II&3955.85 & --&  -- & --&  -- & --&  -- & --&  -- & --&  --\\
  N II&3995.00 & --&  -- & 43&6.97 & --&  -- & 45&6.94 & $<$60&$<$7.29\\
  N II&4447.03 & --&  -- & --&  -- & --&  -- & --&  -- & --&  --\\
  N II&4601.48 & --&  -- & --&  -- & --&  -- & --&  -- & --&  --\\
\end{tabular}
\end{table*}

\setcounter{table}{0}
\clearpage
\begin{table*}
\caption[]{-continued.}
\begin{tabular}{lccccccccccc}
\hline \hline
Species&Wavelength&\multicolumn{2}{c}{NGC2004-101}
&\multicolumn{2}{c}{NGC2004-103}&\multicolumn{2}{c}{NGC2004-104}
&\multicolumn{2}{c}{NGC2004-105}&\multicolumn{2}{c}{NGC2004-106}\\
       &          &EW&Abund&EW&Abund&EW&Abund&EW&Abund&EW&Abund\\
\hline \\
  C II&4267.00 &133&7.65 &143&7.71 &119&7.97 &125&8.00 &153&7.79\\
  N II&3955.85 & --&  -- & --&  -- & --&  -- & --&  -- & --&  --\\
  N II&3995.00 & 85&7.47 & 24&6.73 & $<$60&$<$7.28 & $<$50&$<$7.11 & 48&7.01\\
  N II&4447.03 & --&  -- & --&  -- & --&  -- & --&  -- & --&  --\\
  N II&4601.48 & --&  -- & --&  -- & --&  -- & --&  -- & --&  --\\
\end{tabular}
\end{table*}

\setcounter{table}{0}
\clearpage
\begin{table*}
\caption[]{-continued.}
\begin{tabular}{lccccccccccc}
\hline \hline
Species&Wavelength&\multicolumn{2}{c}{NGC2004-107}
&\multicolumn{2}{c}{NGC2004-108}&\multicolumn{2}{c}{NGC2004-109}
&\multicolumn{2}{c}{NGC2004-110}&\multicolumn{2}{c}{NGC2004-111}\\
       &          &EW&Abund&EW&Abund&EW&Abund&EW&Abund&EW&Abund\\
\hline \\
  C II&4267.00 & 87&7.94 &103&7.56 &163&7.76 &128&7.62 &140&7.56\\
  N II&3955.85 & --&  -- & --&  -- & --&  -- & --&  -- & --&  --\\
  N II&3995.00 & $<$34&$<$7.12 & 36&6.94 & 29&6.82 & 42&6.88 & --&  --\\
  N II&4447.03 & --&  -- & 13&6.91 & --&  -- & --&  -- & --&  --\\
  N II&4601.48 & --&  -- & --&  -- & --&  -- & --&  -- & --&  --\\
\end{tabular}
\end{table*}

\setcounter{table}{0}
\clearpage
\begin{table*}
\caption[]{-continued.}
\begin{tabular}{lccccccccccc}
\hline \hline
Species&Wavelength&\multicolumn{2}{c}{NGC2004-112}
&\multicolumn{2}{c}{NGC2004-113}&\multicolumn{2}{c}{NGC2004-114}
&\multicolumn{2}{c}{NGC2004-116}&\multicolumn{2}{c}{NGC2004-117}\\
       &          &EW&Abund&EW&Abund&EW&Abund&EW&Abund&EW&Abund\\
\hline \\
  C II&4267.00 &120&7.57 &148&7.56 &127&7.61 &128&7.61 &126&7.58\\
  N II&3955.85 & --&  -- & --&  -- & --&  -- & --&  -- & --&  --\\
  N II&3995.00 & --&  -- & $<$70&$<$7.29 & 61&7.25 & 38&6.86 & 38&6.86\\
  N II&4447.03 & --&  -- & --&  -- & --&  -- & --&  -- & --&  --\\
  N II&4601.48 & --&  -- & --&  -- & --&  -- & --&  -- & --&  --\\
\end{tabular}
\end{table*}

\setcounter{table}{0}
\clearpage
\begin{table*}
\caption[]{-continued.}
\begin{tabular}{lccc}
\hline \hline
Species&Wavelength&\multicolumn{2}{c}{NGC2004-119}\\
       &          &EW&Abund\\
\hline \\
  C II&4267.00 &123&7.78\\
  N II&3955.85 &  7&6.97\\
  N II&3995.00 & 40&6.91\\
  N II&4447.03 & --&  --\\
  N II&4601.48 & --&  --\\
  N II&4613.86 & --&  --\\
  N II&4630.54 & --&  --\\
  N II&4643.08 & 13&6.96\\
  O II&3912.00 & 50&8.39\\
  O II&3919.30 & --&  --\\
  O II&3945.04 & 36&8.46\\
  O II&3954.36 & 46&8.35\\
  O II&3982.71 & 25&8.23\\
  O II&4069.00 & 88&8.24\\
  O II&4072.15 & 53&8.10\\
  O II&4075.86 & 67&8.20\\
  O II&4078.84 & 21&8.18\\
  O II&4132.80 & 37&8.43\\
  O II&4156.53 & --&  --\\
  O II&4185.44 & --&  --\\
  O II&4317.00 & 63&8.69\\
  O II&4319.63 & 60&8.62\\
  O II&4325.76 & 21&8.54\\
  O II&4349.43 & 63&8.40\\
  O II&4351.00 & 38&8.08\\
  O II&4353.58 & --&  --\\
  O II&4366.00 & 49&8.34\\
  O II&4369.27 & 15&8.27\\
  O II&4395.94 & 43&8.89\\
  O II&4414.90 & 65&8.17\\
  O II&4416.98 & --&  --\\
  O II&4443.01 & --&  --\\
  O II&4452.38 & 33&8.62\\
  O II&4590.97 & 46&8.18\\
  O II&4596.18 & 59&8.47\\
  O II&4638.86 & 52&8.36\\
  O II&4641.81 & 66&8.17\\
  O II&4650.00 & --&  --\\
  O II&4661.63 & 53&8.31\\
  O II&4673.73 & 18&8.43\\
  O II&4676.23 & 46&8.29\\
  O II&4699.00 & 40&8.03\\
  O II&4705.35 & 38&8.08\\
  O II&4710.00 & 21&8.68\\
 Mg II&4481.00 &108&7.17\\
 Si II&4128.05 & 19&7.27\\
 Si II&4130.89 & 22&7.13\\
Si III&4552.62 &100&7.09\\
Si III&4567.84 & 85&7.13\\
Si III&4574.75 & 58&7.21\\
 Si IV&4088.90 & --&  --\\
 Si IV&4116.10 & --&  --\\
 Si IV&4212.00 & --&  --\\
\hline \\
\end{tabular}
\end{table*}

\begin{landscape}
\onecolumn
\pagestyle{empty}
\addtolength{\topmargin}{50mm} 

\setcounter{table}{1}
\begin{longtable}{lcccccccccc}
\caption[]{Atmospheric parameters, projected rotational velocities and chemical
compositions of the sample.} 
\label{t_results} \\ \hline \hline
Star  &  $T_{\rm eff}$ & $\log g$ &  $\xi_{\rm vt}$& $v\sin i$ & Carbon & Nitrogen & Oxygen & Magnesium & Silicon&~~\\
      &  (K)           & (dex)    &  (km\,s$^{-1}$)&(km\,s$^{-1}$)& (dex) & (dex) & (dex) & (dex) & (dex)&~~\\
     \hline \\
\endfirsthead
\hline \hline
Star  &  $T_{\rm eff}$ & $\log g$ &  $\xi_{\rm vt}$& $v\sin i$ & Carbon & Nitrogen & Oxygen & Magnesium & Silicon&~~\\
      &  (K)           & (dex)    &  (km\,s$^{-1}$)&(km\,s$^{-1}$)& (dex) & (dex) & (dex) & (dex) & (dex)&~~\\
     \hline \\
\endhead
\multicolumn{11}{l}{{Continued on Next Page\ldots}} \\ \endfoot
\hline \endlastfoot
NGC6611-001&30300&3.70&10&142& 8.06$\pm$0.29 ( 1)&    7.87$\pm$0.20 ( 4)& 8.36$\pm$0.20 ( 3)& 7.36$\pm$0.25 ( 1)& 7.47$\pm$0.24 ( 3)\\
NGC6611-006&31250&4.00& 8& 20& 8.14$\pm$0.26 ( 1)&    7.58$\pm$0.10 ( 3)& 8.50$\pm$0.12 (28)& 7.36$\pm$0.21 ( 1)& 7.41$\pm$0.20 ( 3)\\
NGC6611-012&27200&3.90& 5& 95& 8.19$\pm$0.24 ( 1)&    7.50$\pm$0.23 ( 1)& 8.55$\pm$0.14 (23)& 7.26$\pm$0.22 ( 1)& 7.45$\pm$0.30 ( 3)\\
NGC6611-020&27000&4.10& 2&220& 7.98$\pm$0.28 ( 1)&    7.88$\pm$0.22 ( 2)& 8.41$\pm$0.15 ( 6)& 7.60$\pm$0.27 ( 1)& 7.47$\pm$0.28 ( 3)\\
NGC6611-021&26250&4.25& 0& 30& 7.92$\pm$0.22 ( 1)&    7.51$\pm$0.12 ( 3)& 8.60$\pm$0.16 (24)& 7.24$\pm$0.22 ( 1)& 7.40$\pm$0.31 ( 3)\\
NGC6611-025&25000&4.10& 0& 95& 8.19$\pm$0.33 ( 1)&    7.84$\pm$0.16 ( 4)& 8.67$\pm$0.27 (14)& 7.47$\pm$0.34 ( 1)& 7.47$\pm$0.32 ( 3)\\
NGC6611-030&22500&4.15& 1& 10& 8.35$\pm$0.21 ( 1)&    7.69$\pm$0.16 ( 5)& 8.57$\pm$0.23 (29)& 7.32$\pm$0.24 ( 1)& 7.46$\pm$0.28 ( 3)\\
NGC6611-032&22500&4.15&--& 70& 8.32$\pm$0.24 ( 1)&    7.73$\pm$0.22 ( 5)& 8.46$\pm$0.32 (13)&                 --& 7.28$\pm$0.33 ( 3)\\
\end{longtable}  
\begin{itemize}
\item {The values in brackets indicate the number of lines
observed for each species. The atmospheric parameters and rotational velocities
are taken from Papers III and IV. Alternative stellar identifiers can be
found in Papers I and II. Abundances are presented on the scale 
12+$\log$[X/H]. A correction has been applied to the carbon abundances.}
\end{itemize}
\end{landscape}

\end{document}